\documentclass[useAMS,usenatbib]{mn2e}
\usepackage{epsfig}
\usepackage{amsmath}
\usepackage{natbib}
\usepackage{longtable}
\usepackage{graphicx}
\usepackage{amssymb}

\def\mnras {MNRAS}

\title[I. The nature of assembly bias]
  {The nature of assembly bias - I. 
Clues from a $\Lambda$CDM cosmology}
\author[I. Lacerna and N. Padilla]
  {Ivan Lacerna$^{1}$ and Nelson Padilla$^{1}$\\
    $^1$Departamento de Astronom\'ia y Astrof\'isica, Pontificia Universidad Cat\'olica de Chile, 
V. Mackenna 4860, Santiago 22, Chile.}
\date{}

\pagerange{\pageref{firstpage}--\pageref{lastpage}} \pubyear{000}

\begin{document}

\label{firstpage}

\maketitle

\begin{abstract}
We present a new proxy for the overdensity peak height for which the
large-scale clustering of haloes of a given mass does \emph{not} vary
significantly with the assembly history.
The peak height, usually taken to be well represented by the virial mass,
can instead be approximated by the mass inside spheres of
different radii, which in some cases can be larger than the virial
radius and therefore 
include mass outside the individual host halo.
The sphere radii are defined as
$r = a \textrm{ $\delta_t$} + b\textrm{ log$_{10}$}\left(M_{vir}/M_{nl}\right) \textrm{,}$
where $\delta_t$ is the age relative to the typical age of galaxies hosted by 
haloes with virial mass $M_{vir}$, $M_{nl}$ is the non-linear 
mass, and $a=0.2$ and $b=-0.02$ are the free parameters adjusted to 
trace the assembly bias effect. 
Note that $r$ depends on both halo mass and age.
In this new approach, some of the objects which were initially
considered low-mass peaks (i.e. which 
had low virial masses) belong to regions with
higher overdensities. 
At large scales, i.e. in the two-halo regime, this model properly  
recovers the simple prescription where the bias responds to the height 
of the mass
peak alone, in contrast to the usual definition (virial mass) that
shows a strong dependence
on additional halo properties such as formation time.
The dependence on the age 
in the one-halo term is also remarkably reduced 
with the new definition.
The population of galaxies whose ``peak height'' changes
with this new definition consists mainly of old stellar populations and are
preferentially hosted by low-mass haloes located near more massive objects. 
The latter is in agreement with
recent results which indicate  
that old, low-mass haloes would suffer
truncation of mass accretion by nearby larger haloes or simply due to the high density
of their surroundings, thus showing an assembly
bias effect.
The change in mass is small enough that the Sheth et al. (2001) mass function
is still a good fit to the resulting distribution of new masses.
\end{abstract}
\begin{keywords}
cosmology: large-scale structure of Universe -
cosmology: theory -\\
cosmology: dark matter -
galaxies: statistics - 
galaxies: formation
\end{keywords}
\section{Introduction}

Many recent models of galaxy
formation assume that galaxy properties
are determined by the haloes in which
they form and not by the surrounding larger-scale environment 
(e.g. Kauffmann et al. 1997; Berlind et al. 2003; Yang et al. 2003;
Baugh et al. 2005).
In this picture, the galaxy population in a halo of a given mass is independent of where 
the halo is located.
This is justified by the standard description of structure formation,
namely the extended Press-Schechter theory 
(EPS, Bond et al. 1991; Lacey \& Cole 1993; Mo \& White 1996), 
which was in 
turn based on both linear growth theory of density perturbations of
an initial Gaussian random fluctuation field
\citep{PS74}  
and the non-linear spherical collapse model.
Furthermore, simulation results as recent as Percival et al. (2003) 
indicated that 
the halo clustering should only 
depend on the mass.\footnote{Although these authors 
mention that a systematic difference
between the clustering  of the set of all haloes of a 
given mass and any of their subsamples could be hidden within the noise
at a level below 20 per cent in the bias.}

However, a few years ago, it was shown that galaxy properties 
such as the
star formation rate
(Gomez et al. 2003; Balogh et al. 2004; Ceccarelli et al. 2008; and
Padilla, Lambas, 
Gonzalez 2010 in observations; 
Gonzalez \& Padilla 2009 in numerical simulations) 
and colours \citep{Rob-Nelson09} 
depend on the large-scale structure. 
Gomez et al. (2003) found that, for a sample of galaxies
in groups and clusters from the Sloan Digital Sky Survey (SDSS),
the star formation rate decreases, compared with the field population,  
starting at $\sim$ 4 virial radii toward the cluster centre.
Gonzalez \& Padilla (2009) used a semi-analytic model of galaxy formation
and found that the fraction of red galaxies
diminishes for galaxies 
farther away from clusters 
(or closer to voids) in 
environments with the same local density.
These results support the view that galaxy populations
also depend on the larger-scale environment, both in models and observations.

Regarding the fact that haloes of the same mass should essentially exhibit 
the same properties,
Gao et al. (2005, hereafter G05) measured that the large-scale clustering 
of haloes of a given mass depends 
strongly on the formation time, for halo masses
M $\le$ 6$\times$10$^{12}$ $h^{-1}$ M$_{\odot}$.
This study, based on $N$-body simulations, showed that 
haloes assembled at high redshift are more strongly correlated than those
of the same mass that assembled recently.  This effect, 
which is not expected from the 
excursion set theory, was termed ``assembly bias,'' which consists in that the 
large-scale clustering of haloes of a given mass varies significantly with their assembly history
\citep{Gao-White07}. 

On the observational side of the assembly bias, Wang et al. (2008) 
found that groups selected from the SDSS with red central galaxies are
more strongly clustered than groups of the same mass but with blue centrals,
being this effect much more important for less massive groups.
In addition to the clustering 
amplitude, Zapata et al. (2009) found that galaxy groups of similar mass and
different
assembly histories show differences in their galaxy population, for example 
in the 
fraction of red galaxies. 
Furthermore, Cooper et al. (2010) 
studied the relationship between the local environment
and properties of galaxies in the red sequence. 
After removing the dependence of the average
overdensity on colour and stellar mass, they still found
a strong dependence on the luminosity-weighted stellar age. Galaxies with older 
stellar populations occupy regions of higher overdensities compared
to younger galaxies of similar colours or stellar masses. 
The latter results show that the concept of assembly bias could be applicable
to galaxies in addition to dark matter haloes, 
and would then affect the physics of galaxy formation.

Other halo properties such as concentration, number of subhaloes,
subhalo mass function, shape, halo spin, major merger rate, 
triaxiality, shape of the velocity ellipsoid, and velocity anisotropy
at a given mass 
show an assembly-type bias effect in cosmological $N$-body simulations
\cite{Wechsler06,Zhu06,Croton07,Bett07,Gao-White07,HT10,FW10}.

The reasons for this assembly bias are not yet fully understood. 
EPS assumes no such environmental 
dependence. At a fixed mass, the Markovian nature of the random walk
trajectories of perturbations smoothed at higher resolution,
which characterise a halo,
is assumed to be independent of the environment encoded in random walks
at lower resolution.
Thus, halo properties 
should not be related with the external environment in haloes of equal mass.
These random walks are obtained using  
a top-hat Fourier-space window function 
to smooth (or to average) the density fluctuations;
this filter in $k$-space allows 
to obtain an analytic expression of the 
halo mass function that is equal to the Press-Schechter formula.
There have been attempts to modify this window function
to consider an environmental dependence. Zentner (2007) 
combined a Gaussian window function and a variable height of the barrier
for collapsed objects, but found an opposite trend for the
assembly bias at low masses. 

Furthermore, correlations between halo parameters do not simply show the same
clustering behaviour. 
Bett et al. (2007) found that both the most nearly spherical haloes and
those with highest spins are more strongly clustered than the average. 
However, this fact contradicts the correlation between spin and shape, 
where more spherical
haloes have on average a slightly lower spin parameter.
Also, for example, the work by
Croton et al. (2007) showed that there are aspects of the assembly history
which are not encoded in halo concentration or formation redshift 
and which correlate with the large-scale environment. 
One possible explanation was suggested by 
Dalal et al. (2008). They claim that the halo assembly bias is
related to the peak curvature of Gaussian random fields in high-mass haloes,
whereas at the low-mass regime 
the bias arises from a subpopulation of low-mass haloes whose mass 
accretion has ceased. 
These haloes 
could have
been ejected out of nearby massive haloes (Ludlow et al. 2009). 
Wang et al. (2009) found
that these ejected low-mass subhaloes have earlier assembly times
and a much higher bias parameter
than normal (not ejected) haloes of the same mass, 
so that they contribute to the assembly bias.
However, they also found that the assembly bias is not dominated by
this population, indicating that effects of the large-scale environment 
on ``normal'' haloes
is the main source for this bias.

Despite the fact that halo mass
continues 
to be the most important parameter to determine the galaxy 
properties, it is relevant 
to study the assembly bias to gain further insights
on the development of the Large-Scale Structure.  
This is particularly significant when 
galaxies are used to constrain cosmological parameters, as 
shown by Wu et al. (2008) in their study of 
the effects of halo assembly bias on galaxy cluster
surveys.  They used the halo concentration to find 
that upcoming photometric projects such as the Dark Energy Survey
(DES) and the Large Synoptic Survey Telescope (LSST) can 
infer significantly biased cosmological parameters from the observed clustering 
amplitude of galaxy clusters if the assembly bias is not taken into account.

Hester \& Tasitsiomi (2010, hereafter HT10) found an assembly-type bias in 
that the rate of major mergers of haloes of a given mass changes
with the local environment.  They proposed  a dynamical explanation for this effect, 
particularly for high densities, based on both tidal stripping,
responsible for the decrease in the major merger rate of galaxy-like haloes, 
and interactions between bound haloes in the outskirts of groups, which are
related with the increase in the merger rate in group-like haloes. 
This plausible explanation applies on scales of out to $\sim$ 250 kpc.

We suggest that, if the initial peak did not collapse completely onto haloes,
their mass will not be an appropriate proxy for the peak height. 
They will present old ages, which would not be the case if the peak finished 
its collapse onto the halo (it would look younger dynamically). 
Therefore, the scale out to which we need to extend the inclusion of mass 
for the peaks
could be as large as or even larger than the scale proposed by HT10 since, 
at low $z$, the initial overdensity may be still spread 
in larger areas around the current collapsed halo. 
Wang et al. (2007) mention a similar idea in that old, low-mass haloes
were part of higher peaks in the initial density field than what is 
revealed by their present-day virial mass.
By means of a semi-analytic model, we will show that, 
at the present time, the assembly bias may well be related with the
infall region of a halo for scales \mbox{80 kpc $<$ r/$h^{-1} <$ 1.5 Mpc,}
a range where the one-halo clustering amplitude between populations
of the same mass but different ages differs strongly 
(see Section \ref{section_two-point}).

The aim of this work is to understand the origin of the assembly bias
and its role in the development of the large-scale structure and on
the galaxy population, beyond the halo mass dependence.
In order to reach this goal, we will study this effect on 
the semi-analytic galaxies of the Lagos, Cora, $\&$ Padilla (2008) model.
This will allow us to compare our results with those obtained from 
observational data in future work, 
so as to provide another test for the $\Lambda$CDM model of the Universe.
Also, we will show that it is fundamental to include the global effect from 
large scales 
on the peak height estimate. The results obtained with this proxy of the 
peak height will be compared with those obtained from the virial mass of host haloes 
by means of the spatial two-point correlation function and infall velocity profiles
for galaxy samples of different relative ages. This new definition of an overdensity 
peak height will not be subject to the assembly bias seen at large separations,
thus objects of the same mass but different ages will show essentially the same clustering 
in the two-halo regime.  
Given that the assembly bias has also been detected separating samples
according to several other parameters than the halo age, in subsequent papers 
we will also investigate its
prevalence when studying galaxies and haloes of different
concentrations, number of satellites, sphericity, and whether our proposed
explanation for the assembly bias also responds when using observational data.

The outline of this paper is as follows.
In Section \ref{sec_data}, we introduce our simulation.
We then perform the statistics of density fields for the simulation
in Section \ref{section_statistics}
to compare our results with those from other authors that show the 
assembly bias effect. The redefinition of the overdensity peak height
by using the two-point correlation function and the infall velocity profile
is developed in Section \ref{sec_redefinition_full}.
The nature of the objects that are being considered with this redefinition 
are shown in Section \ref{sec_freq}. Finally, we discuss our results in
Section \ref{conclusiones}.  The cosmology used here is $\Omega_{tot}$ = 1, $\Omega_m$ = 0.28,
$\Omega_\Lambda$ = 0.72, $\sigma_8$ = 0.9, $h$ = 0.72, unless otherwise indicated.

\section{Data}
\label{sec_data}

We use the SAG2 model by Lagos, Cora, $\&$ Padilla (2008; 
see also Lagos, Padilla, \& Cora 2009), 
which combines a cosmological
$N$-body simulation of the concordance $\Lambda$CDM universe
and a semi-analytic model of galaxy formation.
The numerical simulation consists of a periodic box of 60 $h^{-1}$ Mpc 
on a side 
that contains 256$^3$ dark matter particles with a mass resolution of 
$\sim$ 10$^9$ $h^{-1}$ M$_{\odot}$. The galaxy population in the 
semi-analytic model is generated using the merger histories of 
dark matter haloes. One of the main features of this model is the 
implementation of the Active Galactic Nuclei (AGN) feedback, which 
reduces star formation by quenching the gas cooling process, 
an important effect on massive haloes 
at low redshifts.

One of the most important parameters throughout this work is the age.
We will use it to study the assembly bias effect  
for galaxies of a wide range of luminosities and, also, 
for dark matter haloes of a wide range in mass.

For galaxies, we will use the mass-weighted stellar age or, simply, 
stellar age defined as 

\begin{equation}
t = t_0 - \frac{ \sum {t_i \Delta t_i \dot{M}_{star}} } { \sum \Delta t_i  \dot{M}_{star} } ,
\end{equation}
\\
where $t_0$ is the age of the Universe today, $t_i$ is the time corresponding to the 
i$^{th}$ output of the simulation, 
and $\dot{M}_{star}$ is the star formation rate 
calculated using the stellar mass $\Delta M_{star}$ accreted in a time step
$\Delta t_i$.  We use this parameter, the stellar age, as it 
can be directly obtained from observational data
\cite{K03,Gallazzi05}.

On the other hand, the formation redshift of a dark matter (DM) halo is 
defined as the redshift when it assembled 50 per cent of its final mass at $z = 0$.
It is important to mention that the assembly bias has been detected
by using this definition of age for DM haloes, and thus it will be used in this work. 
There are other definitions that show a weaker or absent dependence of halo clustering 
on the halo formation time, as was shown by Li et al. (2008). 
\newline

\subsection{Age parameter}
\label{section_age_parameter}
 
\begin{figure}
\leavevmode \epsfysize=8.9cm \epsfbox{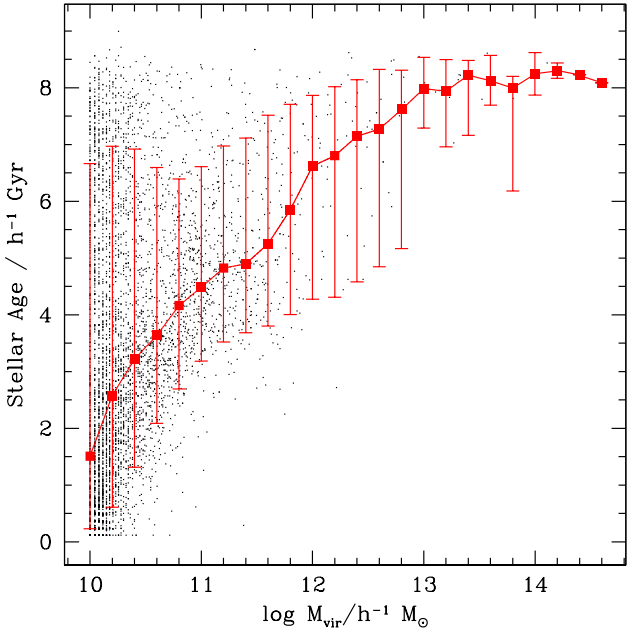} 
\caption{Stellar age as a function of the virial mass 
(logarithms are base 10 throughout) for the 
galaxies in the simulation.
Due to the large number of available galaxies, 
only 5,000 of them, randomly chosen, are 
plotted as points. 
Red squares are the medians for each mass bin. 
Error bars correspond to the 
10 and 90 percentiles of the stellar age distribution. 
The median stellar population of low-mass dark matter haloes is younger 
than that of the massive DM haloes.  
}
\label{Mvir}
\end{figure}

\begin{figure}
\leavevmode \epsfysize=8.9cm \epsfbox{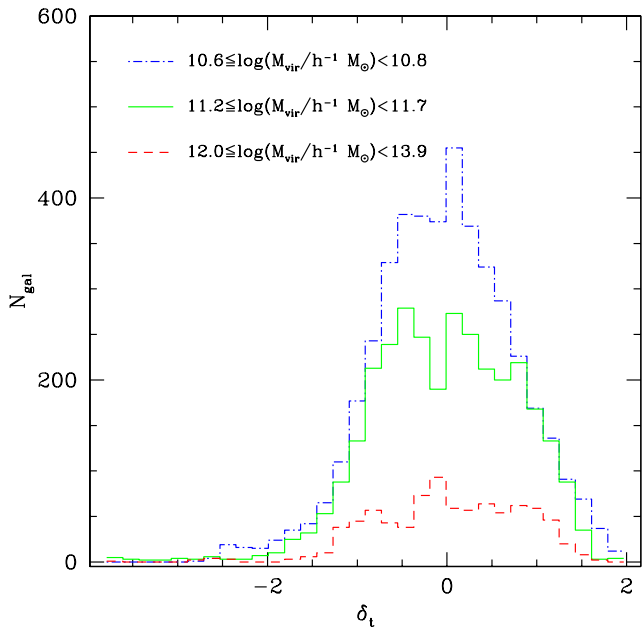} 
\caption{Histograms of the $\delta_t$ parameter for the different mass ranges
indicated in the figure key. 
The three distributions are similar and cover the
full range of $\delta_t$. 
This parameter is independent of the DM halo mass.
}
\label{histo}
\end{figure}

To study the assembly bias, which consists in that
old haloes have a higher clustering than young haloes of the same mass,
it is not convenient to work directly with the 
stellar or halo formation age because they correlate with the mass.
For example, massive dark matter haloes have, on average, 
older stellar populations (Figure \ref{Mvir}).
We need a definition of age which is independent of the mass. This is very important if we want 
to study galaxies in haloes of a wide range of masses.  For example, age maps could show old objects
in regions inhabited only by massive haloes.  Motivated by this problem, the first 
step is to find a proxy for a non-mass-dependent age.
One way to do this consists in using ages relative to the median stellar age as a 
function of the host DM halo mass.    We define the $\delta_t$ dimensionless parameter,

\begin{equation}
\delta_{t_{i}} = \frac{t_{i}- \big<t(M)\big>}{\sigma_t(M)} ,
\label{eq_delta} 
\end{equation}
\\
where, for the i$^{th}$ galaxy, $t_{i}$ is its stellar age, 
$\big<t(M)\big>$ is the
median stellar age as a function of host halo mass (red 
squares connected by the solid line
in Fig. \ref{Mvir}), with M being the virial mass, 
and $\sigma_t(M)$ the dispersion around the median in units of time 
(error bars in Fig. \ref{Mvir}).  In the case of DM haloes, $t_i$ is the formation redshift.
This definition implies that objects with positive (negative) values 
of $\delta_t$ lie above (below) the median stellar age or formation time for a population 
of a given mass.  Then, positive values of $\delta_t$ correspond to older objects, whereas
negative values of $\delta_t$ are related to younger objects. 
The histogram in Figure \ref{histo} shows the distribution 
of the $\delta_t$ parameter for galaxies in different mass bins.
The shape of the distribution of $\delta_t$ is similar among them.
Also, the median host halo mass for $\delta_t <$ 0 and $\delta_t > $ 0 is similar,
$<M>$ $\sim$ 1.7 $\times$ 10$^{10}$ $h^{-1}$ M$_{\odot}$,
confirming that this parameter is independent of the DM halo mass.

\section{Statistics of Density Fields}
\label{section_statistics}

In this section we present the two-point correlation function 
which allows us to calculate the clustering of haloes and galaxies, 
measured directly from the simulation and from theoretical expressions for 
the power spectrum. 

\subsection{The two-point correlation function}
\label{section_two-point}

The correlation function, $\xi(r)$, is a useful quantitative measure
of the spatial clustering. It gives the excess probability for 
finding pairs
of particles at a given separation relative to a Poisson distribution.
The distribution for two points separated by a distance $r$,
with respective volume elements
d$V_1$ and d$V_2$, is given by 

\begin{equation}
dP = n^2[1+\xi(r)]dV_1dV_2 ,
\end{equation}
\\
where $n$ is the average number density of points.

In practice the estimator used, particularly for numerical simulations 
with periodic boundary conditions, is
\begin{displaymath}
DD(r) = RR(r) [1 + \xi(r)], 
\end{displaymath}
and then
\begin{equation}
\xi(r) = \frac{DD(r)}{RR(r)} - 1.
\end{equation}
\\
Here, $DD(r)$ represents the frequency of the data pairs, 
whereas $RR(r)$ corresponds
to the frequency of random pairs, 
defined as

\begin{equation}
RR(r) = N_{sel} N_{tot} \frac{V(r)}{V_{box}} ,
\end{equation}
\\
where $N_{sel}$ is the number of selected objects 
in a given sample, 
$N_{tot}$ is the total number of objects, and
$V(r)$ is the volume in a shell at distance $r$ 
which is normalised by
the volume of the box $V_{box}$ in the simulation.   In the
case of an auto-correlation function, $N_{tot}=N_{sel}$.

The cross-correlation function estimates the clustering amplitude 
between two different data sets.
We will calculate this function for a selected sample
against all the available objects in the simulation because it 
will have 
a higher signal than the correlation 
between the same selected elements, 
i.e. the autocorrelation function 
(e.g. Bornancini et al. 2006).

\subsubsection{Cross-correlation function for haloes}
\label{section_two-point_haloes}

\begin{figure*}
\leavevmode \epsfysize=8.8cm \epsfbox{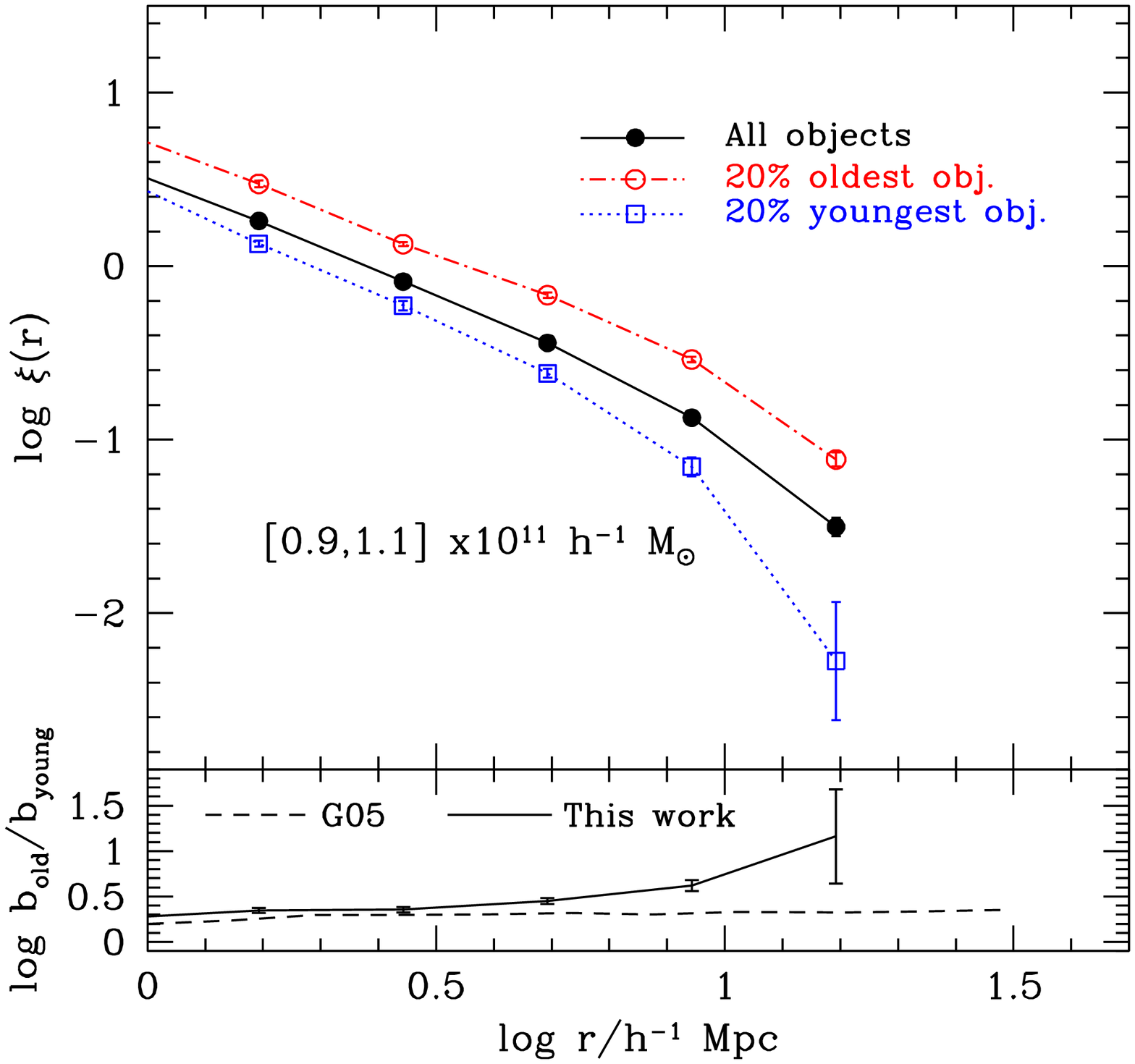} \epsfysize=8.8cm \epsfbox{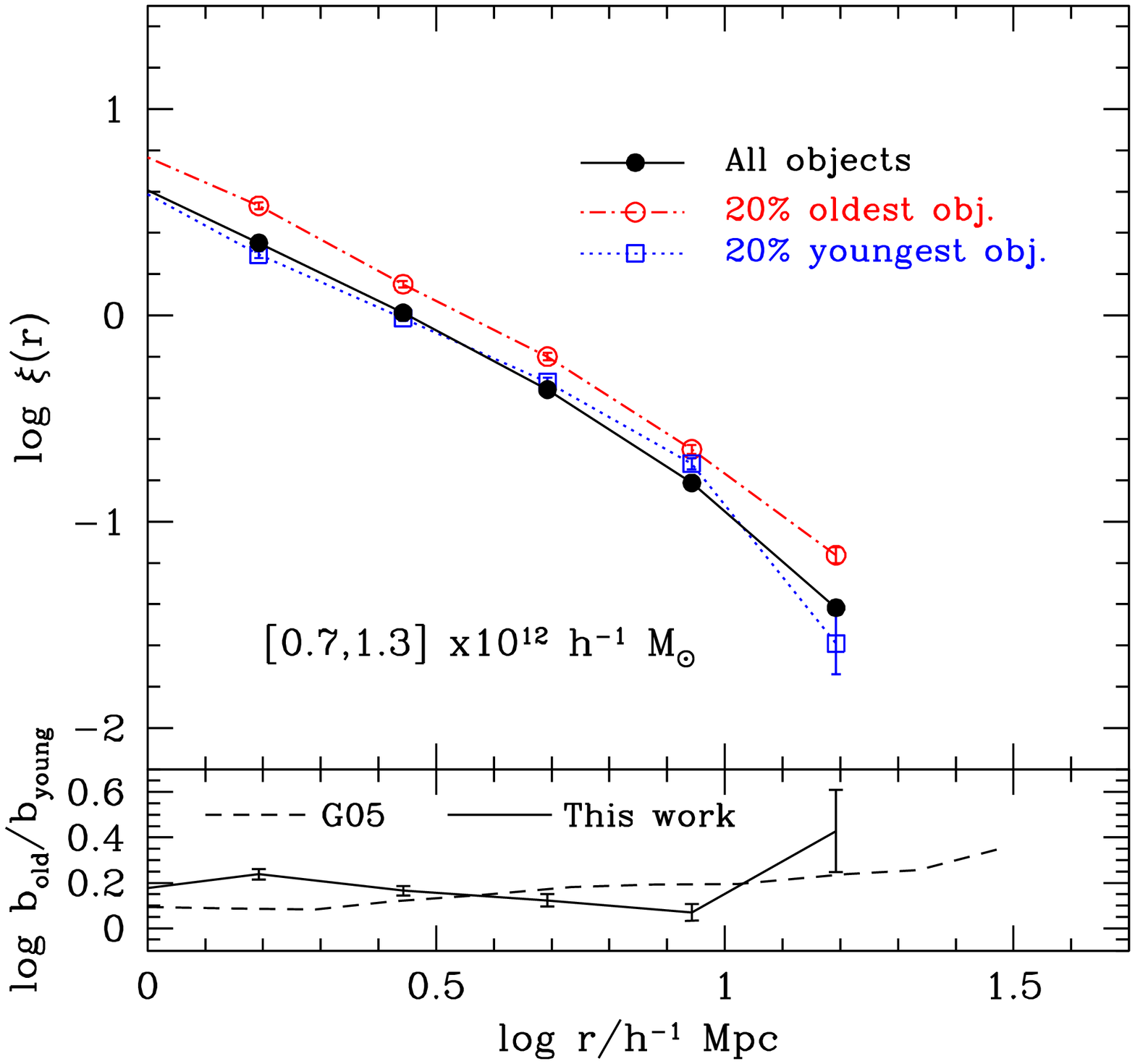} 
\caption{\emph{Main panels (top):} Two-point cross-correlation function for haloes 
from our simulation for two different ranges of mass (left and right panels). 
The old population is represented as dot-dashed red lines, whereas the young one 
appears as dotted blue lines (the full population of haloes are shown as 
solid black lines).
Error bars were calculated using the jackknife method.
\emph{Lower panels:} ratio between the   bias of old and young objects
in our simulation (solid lines) and in G05 (dashed lines).
At large scales, both simulations show a higher clustering for the 
old haloes with respect to the young ones with a remarkable difference 
for the lowest mass bin (left-hand panel).
Our simulation is able to measure the assembly bias effect with
a high statistical significance.
} 
\label{comparacion}
\end{figure*}

In order to test whether our simulation is able to reproduce the observed
signal of assembly bias at large scales found by other authors,
Figure \ref{comparacion} shows the spatial cross-correlation function for haloes of different 
formation times and a given mass against the full population of haloes in the simulation, at $z$ = 0
(top panels).  The two ranges of masses shown in the figure are the same as those used by G05
in two panels of their Fig. 2 which exhibit the assembly bias 
effect.
The result for the 20\% oldest $\delta_t$ haloes in each mass range 
is shown as dot-dashed red lines,
while that for the 20\% youngest haloes is shown as dotted blue lines.
Note that in the panels of their Figure 2, G05 show the autocorrelation function of haloes. 
To compare their estimates of assembly bias with ours, we consider the expression 
found in Mo $\&$ White (1996) for the bias of a given halo sample, $b_H$, on large scales,

\begin{equation}
\xi_{HH}(r, M) = b_H^2(M) \xi_{mm}(r),
\label{eq_halo-matter}
\end{equation}
\\
where $\xi_{HH}$ is the autocorrelation function for haloes and $\xi_{mm}$
is that for the underlying matter, 
which assumes that the halo density field is proportional to the matter
density field times the bias parameter.  
If the population of haloes is separated into old and young subpopulations, 
the ratio between the bias of these samples is, in the G05 case,
\begin{equation} 
\frac{b_{H,old}}{b_{H,young}}= \sqrt{\frac{\xi_{HH,old}}{\xi_{HH,young}}}.
\end{equation}

In our case, we calculate 
the cross-correlation function as

\begin{equation}
\xi_{HH'}(r, M) = b_H(M)b_{H'} (M)\xi_{mm}(r).
\label{eq_halo-matter_cross}
\end{equation}
\\ 
The subscript $H$ refers to the selected haloes
in a mass bin, whereas $H'$ represents all the haloes in our simulation. 
Then, 

\begin{equation} 
\frac{b_{H,old}}{b_{H,young}}= \frac{\xi_{HH',old}}{\xi_{HH',young}} .
\end{equation}
\\
These ratios are shown as dashed and
solid lines in the lower panels of Figure \ref{comparacion}
for the G05 and our simulation, respectively.\footnote{The cosmology 
in G05 was adjusted to that used in this paper.}

As can be seen from these panels, 
both simulations show a 
higher clustering for the old population
than that for the young one, and it can also be seen that
our simulation can reproduce the 
assembly bias effect with an appropriate statistical significance, 
particularly for low-mass haloes.

\subsubsection{Galaxy cross-correlation functions}
\label{section_two-point_galaxies}

\begin{figure*} 
\leavevmode \epsfysize=5.8cm \epsfbox{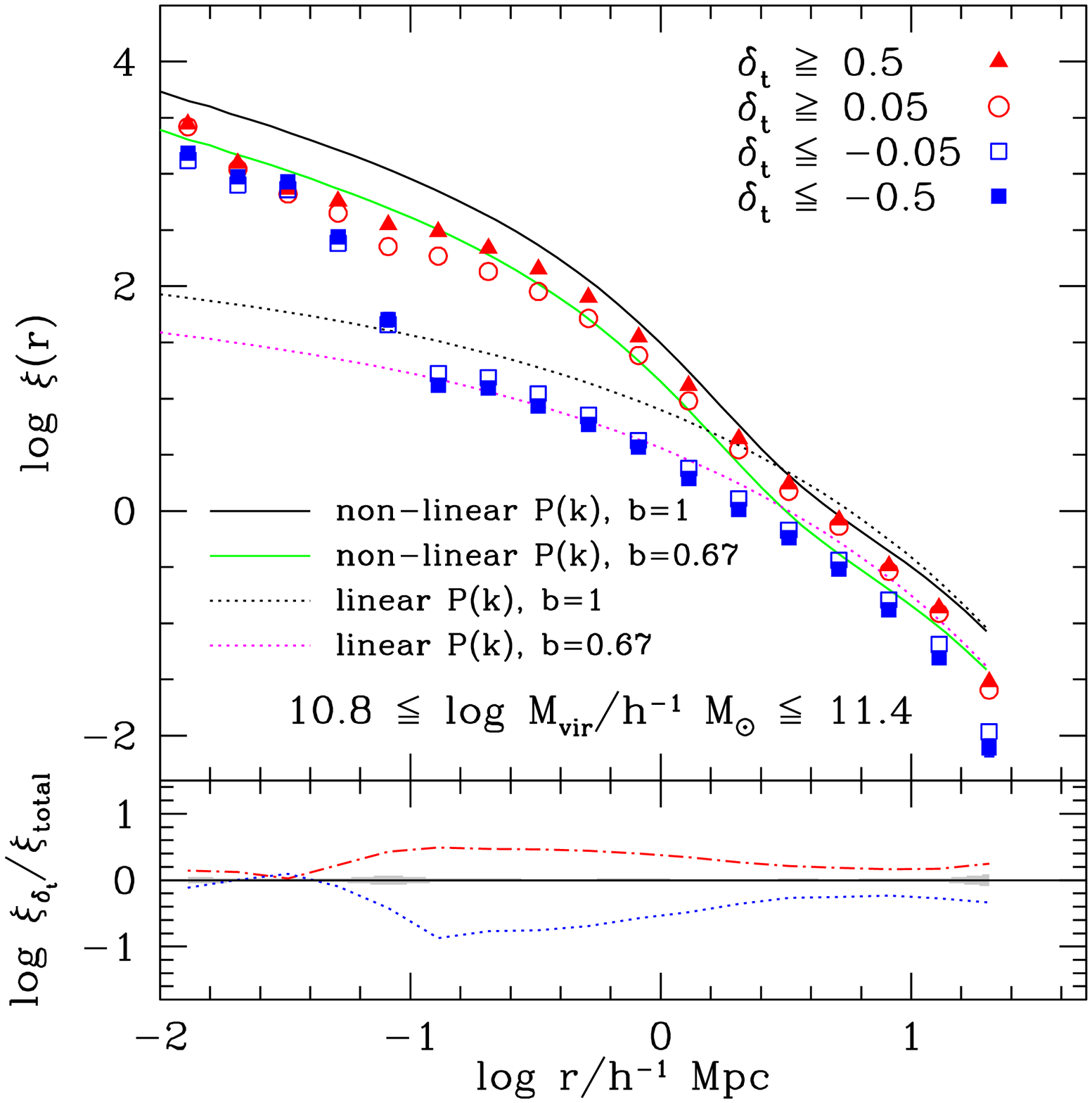} \epsfysize=5.8cm \epsfbox{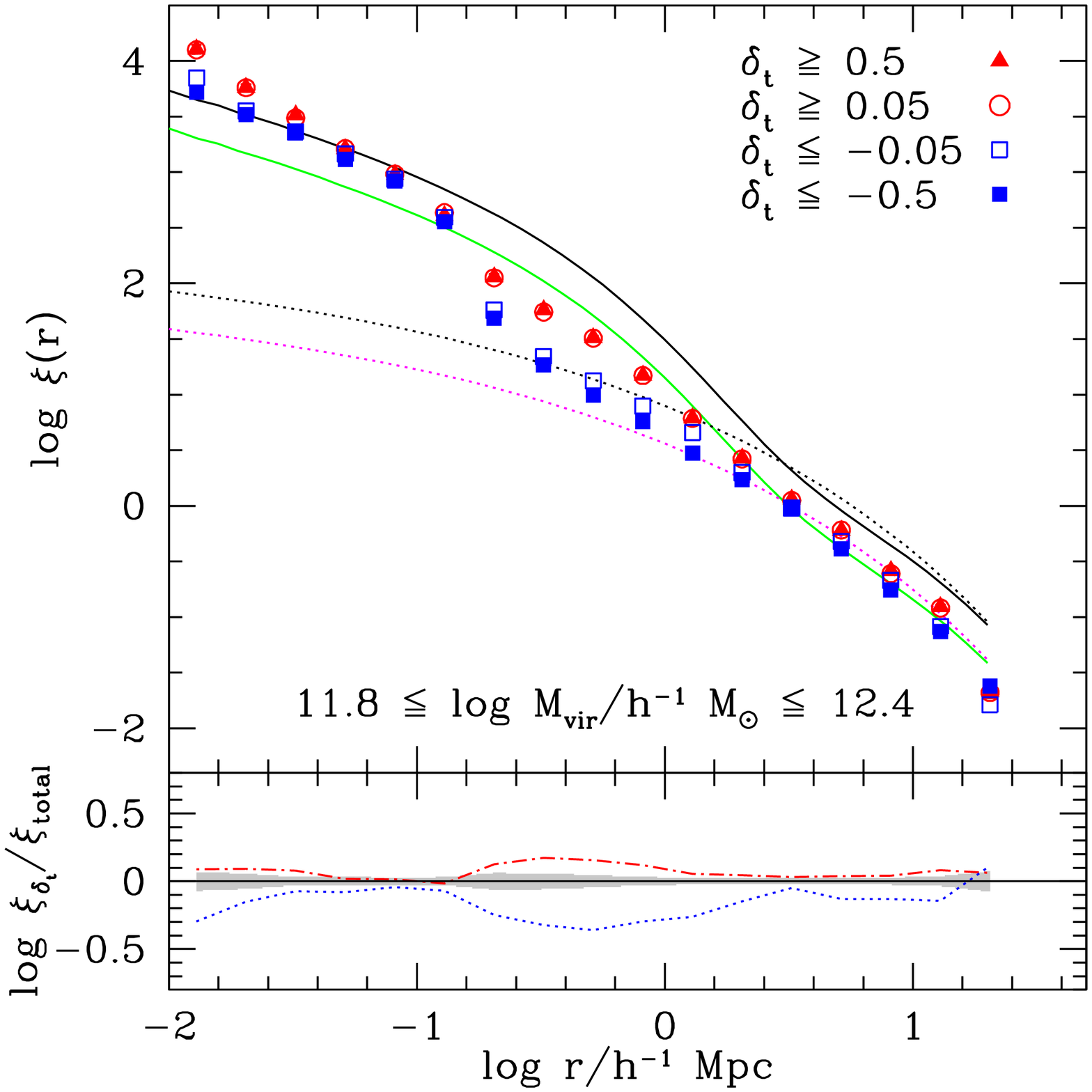}  \epsfysize=5.8cm \epsfbox{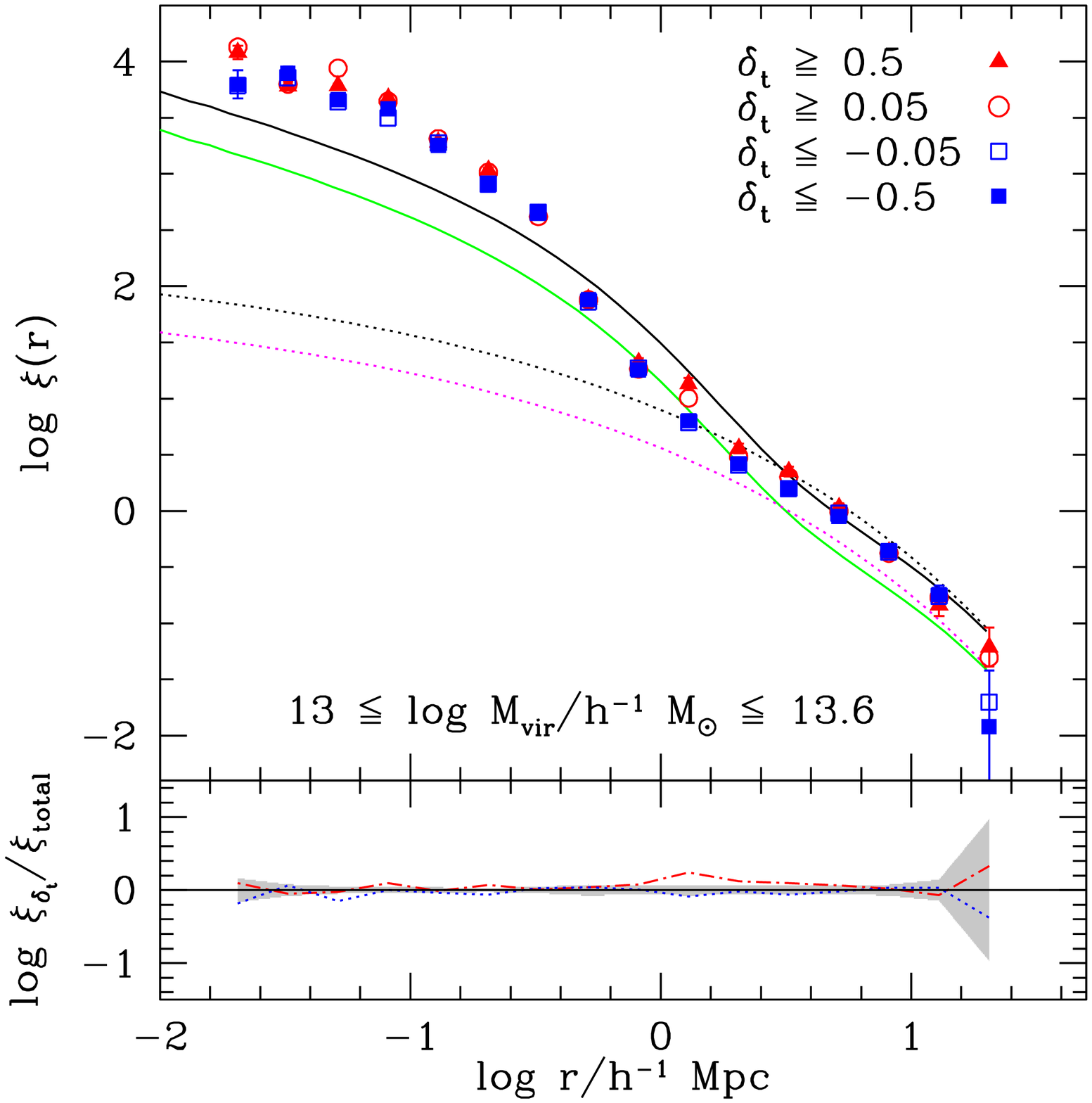}
\bigskip

\leavevmode \epsfysize=5.8cm \epsfbox{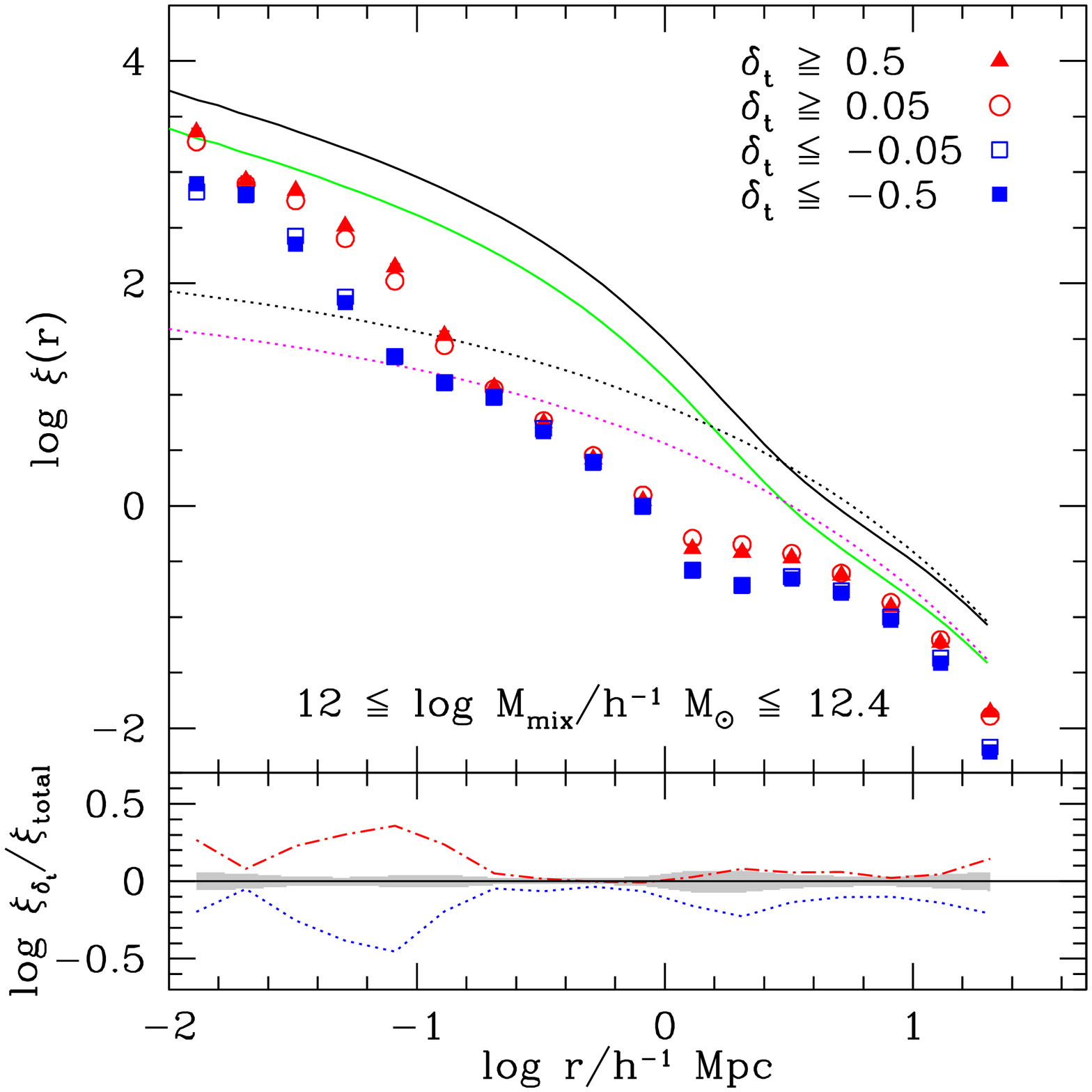} \leavevmode \epsfysize=5.8cm \epsfbox{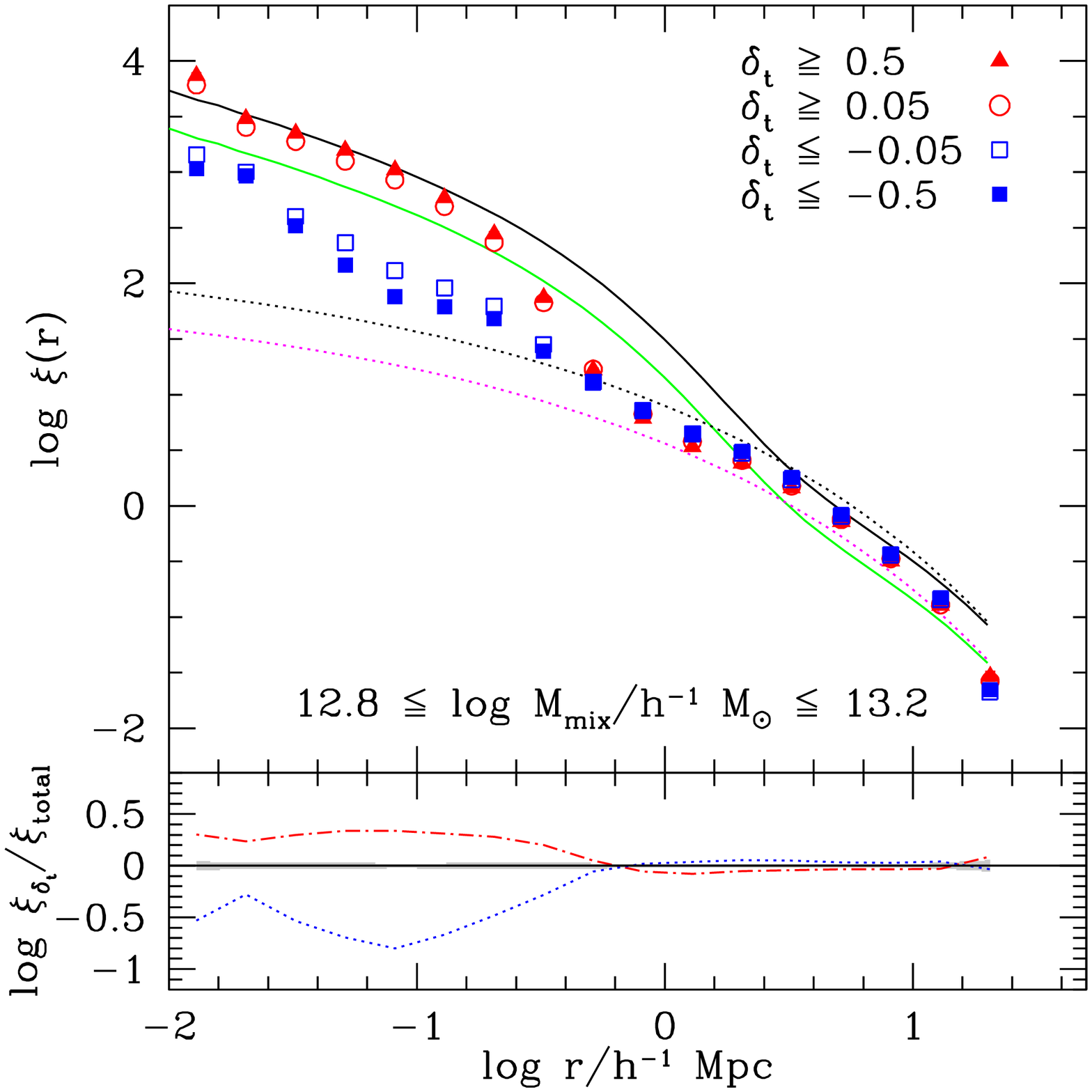} \leavevmode\epsfysize=5.8cm \epsfbox{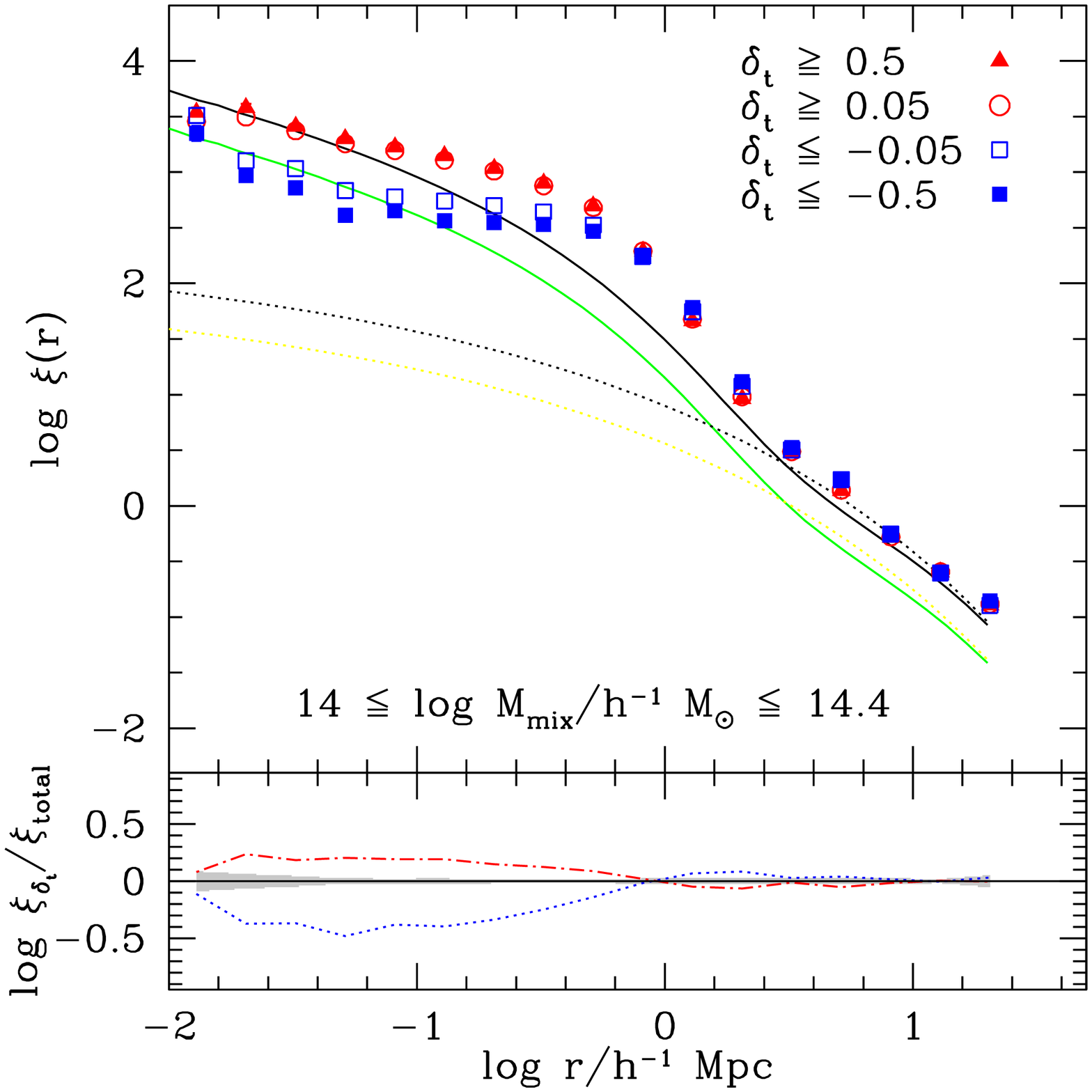}

\caption{Correlation functions for the different mass bins 
indicated in each panel. Old and young galaxies are shown in 
red and blue symbols, respectively. The figure key shows the ranges 
of $\delta_t$ corresponding to the different symbols.
Error bars were calculated using the jackknife method.
The lines repeated in each top box 
are obtained from the 
non-linear and 
linear power spectra, $P(k)$ 
(labels indicated in the top left panel. See details in Section \ref{section_PS}).
The lower box 
in each panel corresponds to the 
ratio between the 
correlation function of the
oldest population (red triangles) and the total population of the 
selected sample and, also, 
between the youngest
population (filled blue squares) and the total one (dot-dashed red and 
dotted blue lines,
respectively).
The error of the ratio between the $\xi$(r) of the oldest  
and youngest objects is shown as a shaded region around the 
unit ratio.
\emph{Top row}: The age definition using the virial mass of the host halo.
Notice the strong difference of almost two orders of magnitude 
between the old and young populations at $r \sim$ 150 $h^{-1}$ kpc 
for the lowest mass bin.
\emph{Bottom row}: Galaxies are selected
according to a tentative new mass measurement, $M_{mix}$ 
(see Section \ref{new_halo}).
}
\label{xi_bines_M}
\end{figure*}

The top row
of Figure \ref{xi_bines_M} shows the spatial 
cross-correlation function 
between galaxy samples
of different relative ages $\delta_t$ 
but equal host halo masses,
and the full population of galaxies in our simulation ($\sim$ 63,000 objects).
Using $\delta_t$, we find an assembly bias effect in our galaxies where
the old population (red filled triangles and open circles) shows 
a higher clustering than the young population (blue open and filled squares), being
this effect much stronger for the low-mass regime. 
The lower box in each panel shows the ratios between
the correlation function of the oldest
population (red triangles) and the total population,
and between the youngest
population (filled blue squares) and the total one, 
as dot-dashed red and dotted blue lines, respectively.
The error of the ratio between 
$\xi$(r) for 
the oldest and 
youngest objects is shown as a shaded region around the value that would be obtained if both
correlation functions were the same (ratio equal to unity).

As can be seen from the lowest-mass bin (top left panel), 
the amplitude of clustering is higher for the old population
than the young one, particularly for scales  
\mbox{80 kpc $<$ r/$h^{-1} <$ 1.5 Mpc} (one-halo term). 
This could indicate 
that their density profiles are different, probably 
those of the young population
being dynamically less internally evolved.
The strong difference in clustering at distances beyond 1 $h^{-1}$ Mpc 
may imply that if the mass in the vicinities 
(surrounding areas or the infall region) 
of haloes were taken into account, 
it would show no dependence on age. 
In other words, as the virial mass of haloes is not good enough as an
overdensity peak height estimator in the simple EPS picture,
this alternative could provide a better estimator for this peak height.
HT10 detected
an assembly-type bias for the dark matter halo major merger rate
using the Millennium Simulation \cite{Springel05}, 
and proposed
a physical mechanism for this effect that, as was mentioned above, extends 
out to $\sim$ 250 kpc.
Owing to the result seen in the top row of Figure \ref{xi_bines_M},
it is possible that  
to explain the assembly bias 
one would need to characterise the peak with mass on scales larger
than the virial radius
(see Section \ref{sec_redefinition_full}),
extending the local 
definition of peak from
within a halo to larger scales
usually regarded as part of the global environment.

Additionally, recent studies have suggested that a population of subhaloes
that were expelled from larger haloes
located beyond three times the virial radius of the main halo could
explain the age dependence of the clustering in the low-mass regime
\cite{Dalal08,Ludlow09}.  
Although Wang et al. (2009)
found that
these low-mass haloes are not the main source for the assembly bias, they claim
that environmental effects at large scales have a very important role
on this issue. 
In this case, the mass of the expelled subhaloes are bad indicators of their
peak height, which could be better represented by the higher mass of a 
larger halo.

Note that our results 
reproduce those found by previous authors where 
the dependence of the clustering on the assembly history is
only detected in low-mass haloes. This indicates that the peak may include
matter around haloes to distances that depend on both halo mass  
and age.

\subsection{Theoretical estimates of $\xi$(r)}
\label{section_PS}

The statistical properties of the density fluctuation field 
can be represented 
by its power
spectrum $P(k)$, or equivalently by its dimensionless 
power spectrum $\Delta^2(k)$,
\begin{equation}
\Delta^2(k) = \frac{k^3}{2\pi^2}P(k),
\end{equation}
\\
which measures the power per logarithmic unit bin in wavenumber $k$.
This spectrum is a direct manifestation of the hierarchical growth of
structures, where small-scale perturbations collapse first 
to grow and later form larger-scale perturbations which will collapse
and form larger objects,
in a non-linear process as time progresses.
Directly related to this evolution are the abundance and 
clustering of galaxy systems and their variations 
as a function of mass and redshift. The Fourier transform of the power spectrum
results in the matter correlation function

\begin{equation}
\xi_{mm}(r) = \int \Delta^2(k) \frac{\sin(kr)}{kr}\frac{dk}{k}  .
\label{eq_xi_mm}
\end{equation}

The Smith et al. (2003) fitting model 
provides a good 
estimate for $\xi_{mm}(r)$ from
the non-linear and the linear power spectra
\cite{Boylan-Kolchin09}. 
In order to do the comparison with the correlation 
function of galaxies in the simulation,  
we use Equations (\ref{eq_halo-matter_cross}) and (\ref{eq_xi_mm}).
The bias parameter $b$ is estimated by using the
fit proposed by Seljak $\&$ Warren (2004),

\begin{eqnarray}
b_0(x = M/M_{nl}) & = & 0.53 + 0.39x^{0.45} + \frac{0.13}{40x + 1} \nonumber\\
& & {}+ 5 \times 10^{-4} x^{1.5},
\end{eqnarray}
\\
with an accuracy on the bias-halo mass relation at the level of 3 per cent
for $b <$ 1.
Here, $M_{nl}$ refers to the non-linear mass, 
defined as the mass within a sphere
for which the rms fluctuation amplitude of the linear field is 1.69 times 
the critical density of the Universe which corresponds to the gravitational 
collapse in the spherical collapse model \cite{GG72}.  
For our 
simulation, we find 
$M_{nl} \sim$ 2.4 $\times 10^{13}$ $h^{-1}$ M$_{\odot}$.

The top boxes 
in each panel of Figure \ref{xi_bines_M} show 
$\xi(r)$ as obtained from the 
non-linear power spectrum (solid black line) with $b$ = 1;
from the non-linear $P(k)$ (lower solid green line) 
with $b$ = 0.6733 (corresponding to the average host halo mass); 
from the linear $P(k)$ with $b$ = 1 (dotted black line); and 
from the linear $P(k)$ (lower dotted magenta line) with  
the bias factor for the average host halo mass.
The biased $\xi(r)$ obtained from the non-linear $P(k)$ (lower solid green line) 
is expected to represent the correlation function 
without the assembly bias for the average host halo mass.
The scale where the linear and non-linear power spectra
start to diverge is around $\sim$ 1.5 $h^{-1}$ Mpc, 
and it is also out to where the correlation function shows the
stronger difference in clustering between the old and young populations
(top left panel of Figure \ref{xi_bines_M}). 
Therefore, we will start studying scales of this tentative size for estimating 
the height of the mass peak 
to see how it affects the galaxy clustering (Section \ref{new_halo}).
Later in this paper we will carry out a $\chi^2$ search for this scale and
its dependence on halo properties,
since the scale may introduce a 
large 
change in the resulting peak mass function.

\section{Redefinition of an overdensity peak height}
\label{sec_redefinition_full}

We propose to extend the 
proxy for peak height
to larger scales
so that it does not show the assembly bias effect.
The scale will  
include the mass of the peak which 
has already collapsed but also, in some cases, 
some of the mass that, due to global environmental effects, 
has not done so yet.
This will be equivalent to a new definition
of ``halo.''
For each galaxy we will consider all the dark matter particles 
within a scale 
that will depend on the host halo mass and its age (see Section \ref{sec_prm}).
The mass contained in this halo, together with the stellar age of each
galaxy,  
will be used to study the large-scale bias.

Throughout this section, two different approaches that characterise
the assembly bias, the two-point correlation function and 
the infall velocity profile,
will be presented and later used to define the overdensity peak height.  
They will allow one to parametrise the  
scale which
will trace the assembly bias at large scales.

\subsection{Using $\xi(r)$ to determine the presence of an assembly-type bias}
\label{new_halo}

In a first attempt, we approximate the 
peak height
for each galaxy considering all the DM particles inside
a radius of 1.5 and 1.7 $h^{-1}$ Mpc for old and young objects, respectively,
motivated by the results of the previous section.
We then repeat the same procedure
described in Section \ref{section_age_parameter}. 
In this case, Equation (\ref{eq_delta}) is 
applied using this new mass definition, $M_{mix}$. 
The results are shown in the bottom row of Figure \ref{xi_bines_M}.
The left column shows the lower mass bin, 
whereas the most massive bin is shown
in the right column.
This peak height definition cannot fully trace the assembly
bias at large scales for each mass bin, 
although it does a better 
job than the virial mass. 
Therefore, a redefinition of halo mass
including larger scales than the virial radius 
could recover the simple prescription where
the bias responds to the height 
of the mass peak alone.
To achieve this goal, 
it thus seems necessary to consider influences beyond the virial radius, 
probably reaching the infall region of haloes.

It is important to point out that the redefinition of mass 
does not affect positions and hence only changes the relative 
age $\delta_t$ of a galaxy.

Since this tentative approximation may be overcorrected as the scale may depend on 
different parameters (e.g. mass, age),
the next section will present an estimator
for the size of the region to use for
this new definition of peak height based also on the velocity profile 
of the infall region. Then, we will combine the constraints obtained from
correlation functions and infall velocities to estimate a parametrised proxy 
for the peak height.

\subsection{Using the Infall Velocity to determine the presence of an 
assembly-type bias}
\label{section_vel_inf}

\begin{figure*}  
\leavevmode \epsfysize=8.7cm \epsfbox{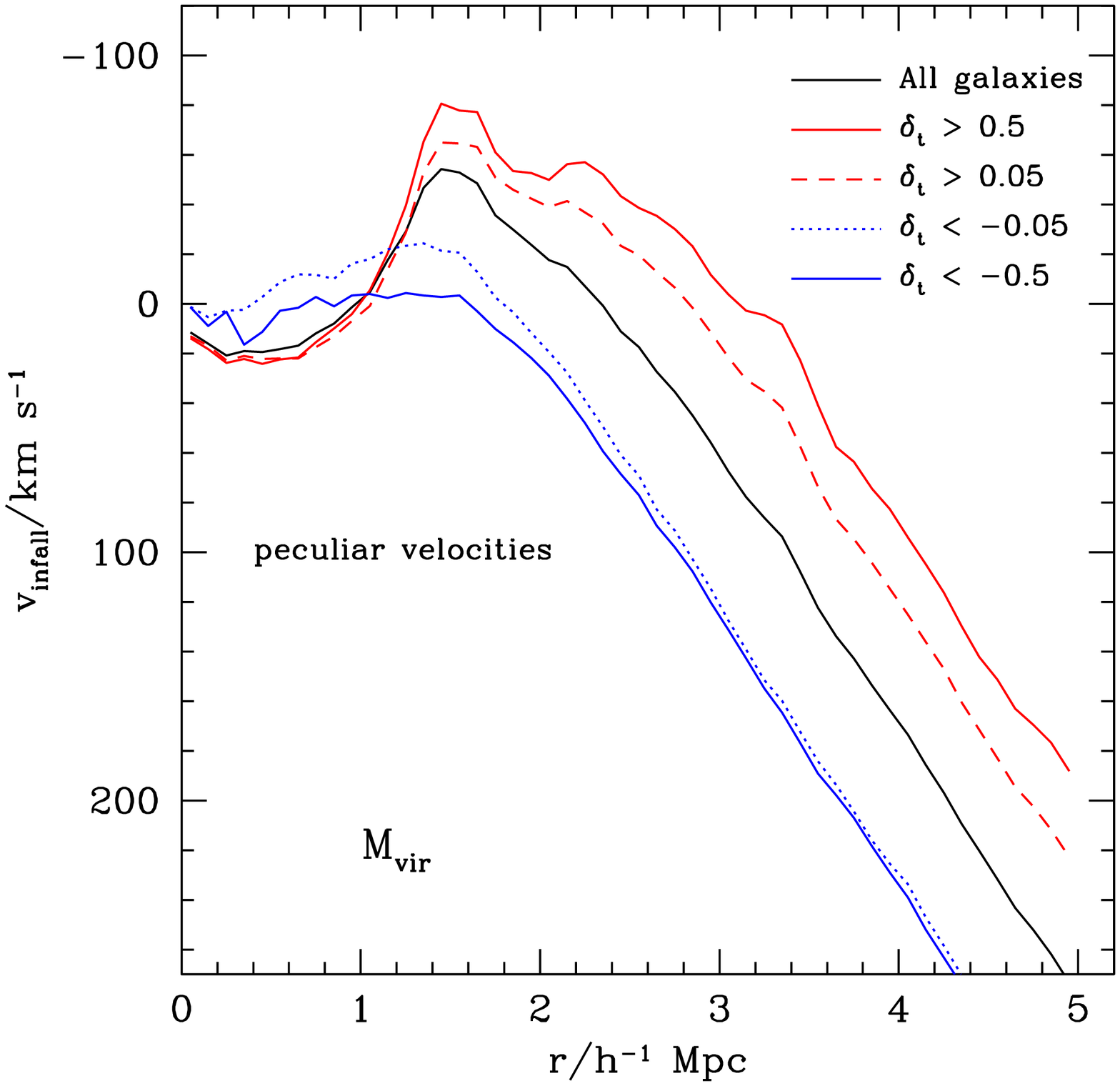} \leavevmode \epsfysize=8.7cm \epsfbox{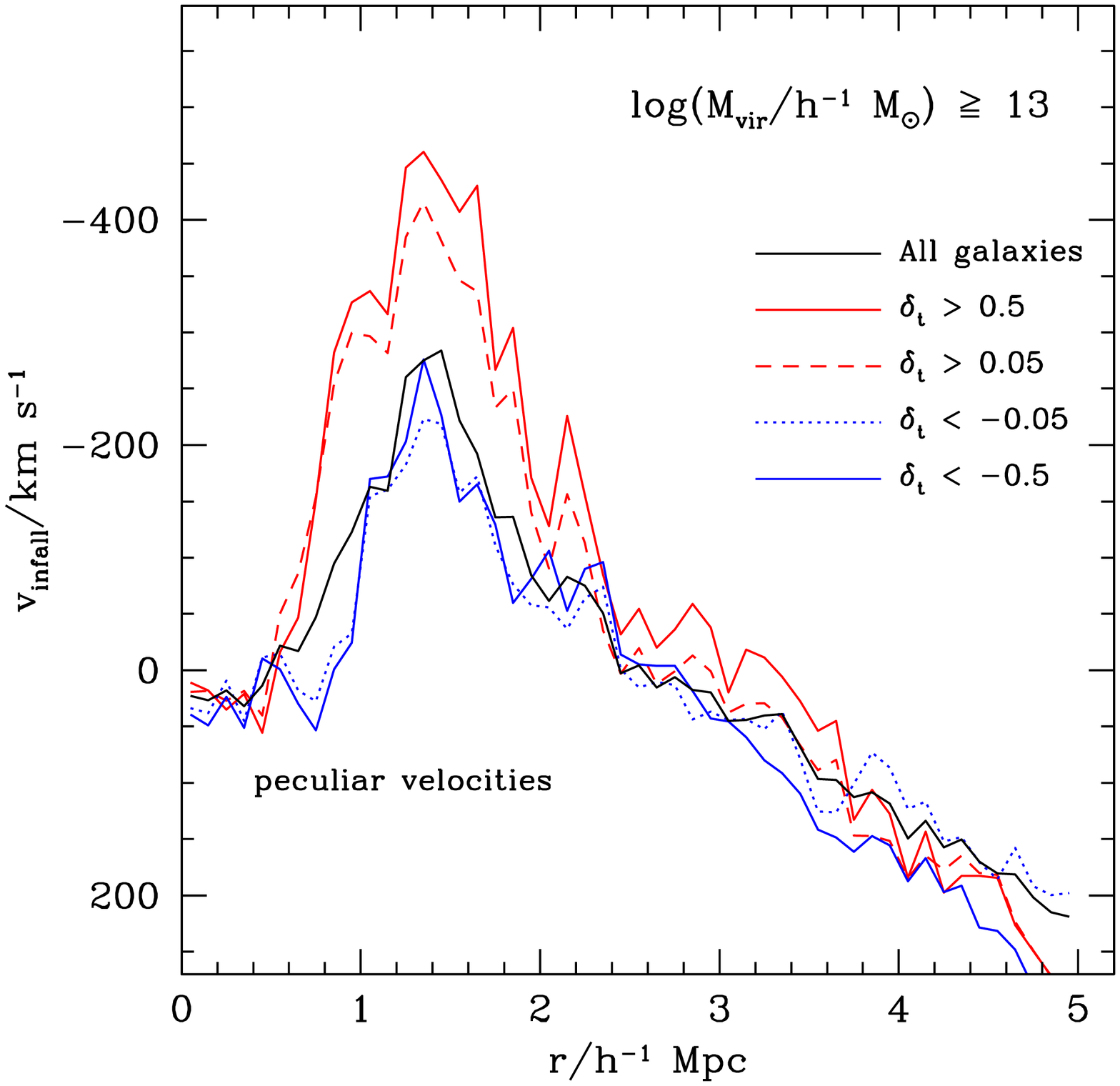}
\caption{Radial velocity profiles around galaxies.
The age parameter $\delta_t$ is given by the virial mass.
The figure keys show the ranges of $\delta_t$ corresponding to the different symbols.
The black solid line 
corresponds to the situation where the galaxies are not split
according to their ages.
\emph{Left:} All haloes.  In this case, 
the infall velocity for 
young galaxies (lower dotted and solid blue lines) does not have as clear a peak 
as that for old objects  (upper solid and dashed red lines).
\emph{Right:} A subsample with high-mass objects, 
$M_{vir} \geq 10^{13}$ $h^{-1}$ M$_{\odot}$. 
The old and young galaxies (red and blue, respectively) 
have an akin profile,  
as both populations show a peak
at $\sim$ 1.3 $h^{-1}$ Mpc. 
This correlates with the smaller difference in clustering
between 
old and young galaxies hosted by high-mass haloes.
}
\label{infall_Mvir}
\end{figure*}

The infall velocity profile $v_{inf}$ around galaxies is another statistic which
is sensitive to the assembly bias.
The $v_{inf}$ values should depend on the initial density fluctuations as well
as the clustering because, at large scales,
the behaviour of haloes (and galaxies)
is dominated by the collapse of these perturbations.

We calculate the radial velocity profile around galaxies, which
can be expressed as

\begin{equation}
v_{r}(r) = v_{n_i}(r) - v_{c_i}, 
\end{equation}
\\
where $v_{c_i}$ is the projected velocity of the central galaxy of the new halo
along the  
direction between this galaxy and its i$^{th}$ 
neighboring galaxy located at a distance $r$, 
whose projected velocity along this 
direction is $v_{n_i}(r)$.
The infall velocity at a distance $r$ around galaxies 
is the average value of $v_{r}(r)$.

The $v_{inf}$ profiles for young and old galaxies 
according to their virial masses are quite different 
(left-hand panel of Fig. \ref{infall_Mvir}). 
While the old galaxies (upper solid and dashed red lines) have a peak in their infall 
velocity distribution 
at 1.5 $h^{-1}$ Mpc, the young population does not
have a clear maximum  (lower dotted and solid blue lines). 
The top of this distribution is rather
flat around 1.3 $h^{-1}$ Mpc. 
However, galaxies located in high mass haloes as those plotted in the top
right panel of Fig. \ref{xi_bines_M} should have similar velocity profiles,
since they 
do not have a strong assembly bias.
Their velocity profiles are shown in 
the right-hand panel of Figure \ref{infall_Mvir}.
Both populations show similar infall velocity behaviours and
a peak of the distribution at $\sim$ 1.3 $h^{-1}$ Mpc.

The aim of the next section is to find the best values of the radius
enclosing the mass of the density peak
as a function of both age and mass,
in order to obtain similar velocity profiles 
and correlation functions 
for galaxies of very different ages but equal masses.

\subsection{Parametrising a new overdensity peak height proxy}
\label{sec_prm}

The previous sections have shown that a new proxy for the peak height 
could better characterise on average the assembly bias effect seen at 
large scales than the proxy given by the virial mass. 
Apart from the difference in clustering between populations of 
different ages but equal mass, the velocity profile could also be used to detect this bias.

We parametrise the radius of each galaxy as a function of both virial mass and 
$\delta_t$.  
We then measure the masses inside spheres 
defined by this radius and calculate their relative ages with respect to 
this mass. 
Finally, a $\chi^2$ statistics between the young and 
old populations of the differences between velocity profiles, 
$\chi^2_{v(r)}$, and correlation functions, $\chi^2_{\xi(r)}$,
will be used to find the best parameter set that traces more accurately the 
assembly bias.

The radius for each galaxy is parametrised as 

\begin{equation}
r = a \textrm{ $\delta_t$} + b\textrm{ log}\left(\frac{M_{vir}}{M_{nl}}\right) \textrm{ ,} 
\label{eq_r_prm}
\end{equation}
\\
where $M_{nl}$ is the non-linear mass defined by Seljak \& Warren 
(2004, see Section \ref{section_PS}), 
log($M_{nl}$/$h^{-1}$ M$_{\odot}$) = 13.38 
for our choice of cosmological parameters.
The free parameters are $a$ and $b$. 
The new peak height proxy will be the mass $M$ enclosed within this radius.
It is assumed that if $r$ is smaller than the virial radius $r_{vir}$
or if $M$ is smaller than the virial mass, 
then $M = M_{vir}$.

Once the new mass contained in this radius 
and its $\delta_t$ are measured,
infall velocities and correlation functions are 
calculated for three bins in mass corresponding to the
first, second, and third terciles of the mass
distribution. This selection produces results for low, medium, and high
masses, respectively.

The $\chi^2$ for the infall velocity field is calculated as

\begin{equation}
\chi^2_{v(r)} = \frac{1}{3} \sum_{i}^3 \left(\frac{1}{n_{dof}}  \sum_r \frac{ \big[v_{neg}(r) - v_{pos}(r) \big] ^2}{\sigma^2_{v(r)}}\right)_i 
\textrm{ .}
\label{eq_chivel}
\end{equation}
\\
The $\chi^2$ value for the $i^{th}$ mass bin
is performed within the range 2.5 $\leq r/h^{-1}$ Mpc $\leq 5$, 
since this interval 
corresponds to the two-halo regime;
$v_{neg}$ is the mean radial velocity around galaxies with
$\delta_t < -0.05$ (young objects) and 
$v_{pos}$ is this same quantity for $\delta_t >$ 0.05 (old galaxies). 
The error is estimated
as $\sigma^2_{v(r)} = \sigma^2_{v_{neg}} + \sigma^2_{v_{pos}}$, 
with the first term being
the error for $v_{neg}$ and the second term 
the error for $v_{pos}$,
calculated as the error of the mean within the interval of interest. 
The symbol $n_{dof}$ denotes the number of degrees of freedom.
The value $\chi^2_{v(r)}$ is the average over the three mass bins. 

Similarly, the reduced $\chi^2$ for the correlation function statistics 
is defined as
\begin{equation}
\chi^2_{\xi(r)} = \frac{1}{3} \sum_{i}^3 \left(\frac{1}{n_{dof}} \sum_r \frac{ \big[N_{neg}(r) - N_{pos}(r) \big]^2}{\sigma^2_{N(r)}}\right)_i 
\textrm{ ,}
\label{eq_chixi}
\end{equation}                                                                          
\\
and is calculated within the range 0.8 $\leq r/h^{-1}$ Mpc $\leq 10$,
mostly in the two-halo term;
$N_{neg}$ is the number of neighbours for young galaxies, 
whereas $N_{pos}$ is 
the same quantity for old objects. 
The number of neighbours is defined as
$N(r) =$ $<N_t(r)> \xi(r)$ $+ <N_t(r)>$, 
where  $<N_t(r)> = N_{pairs}(r)/N_{centres}$ 
is the mean number of 
tracers.\footnote{This relation comes from 
$\xi(r)=\frac{N(r) - <N_t(r)>}{<N_t(r)>}$. Therefore, if the distribution
of neighbours is random, $\xi(r)$ would be equal to zero.}
The error is 
$\sigma^2_{N(r)} = \sigma^2_{N_{neg}} + \sigma^2_{N_{pos}}$, 
where the first term is
the error for $N_{neg}$ and the second for $N_{pos}$,
both calculated as the relative error of the number of neighbours.
We choose this alternative to normalise the reduced $\chi^2$ 
in order to avoid selecting parameters favoured by
large uncertainties that can
induce spuriously good fits.  
The value $\chi^2_{\xi(r)}$ is the average over the three mass bins. 

\begin{figure*}
\leavevmode \epsfysize=5.8cm \epsfbox{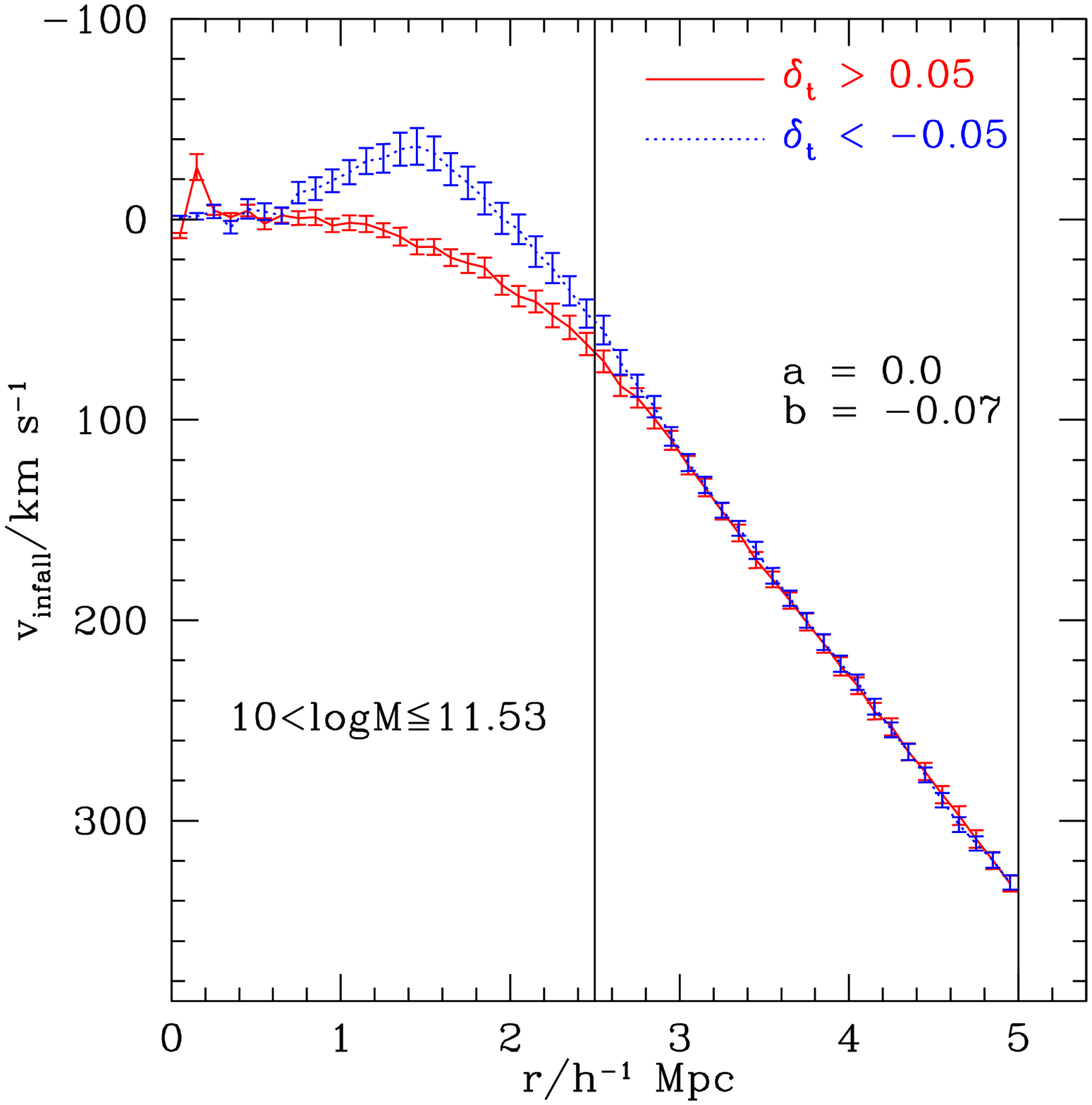} \epsfysize=5.8cm \epsfbox{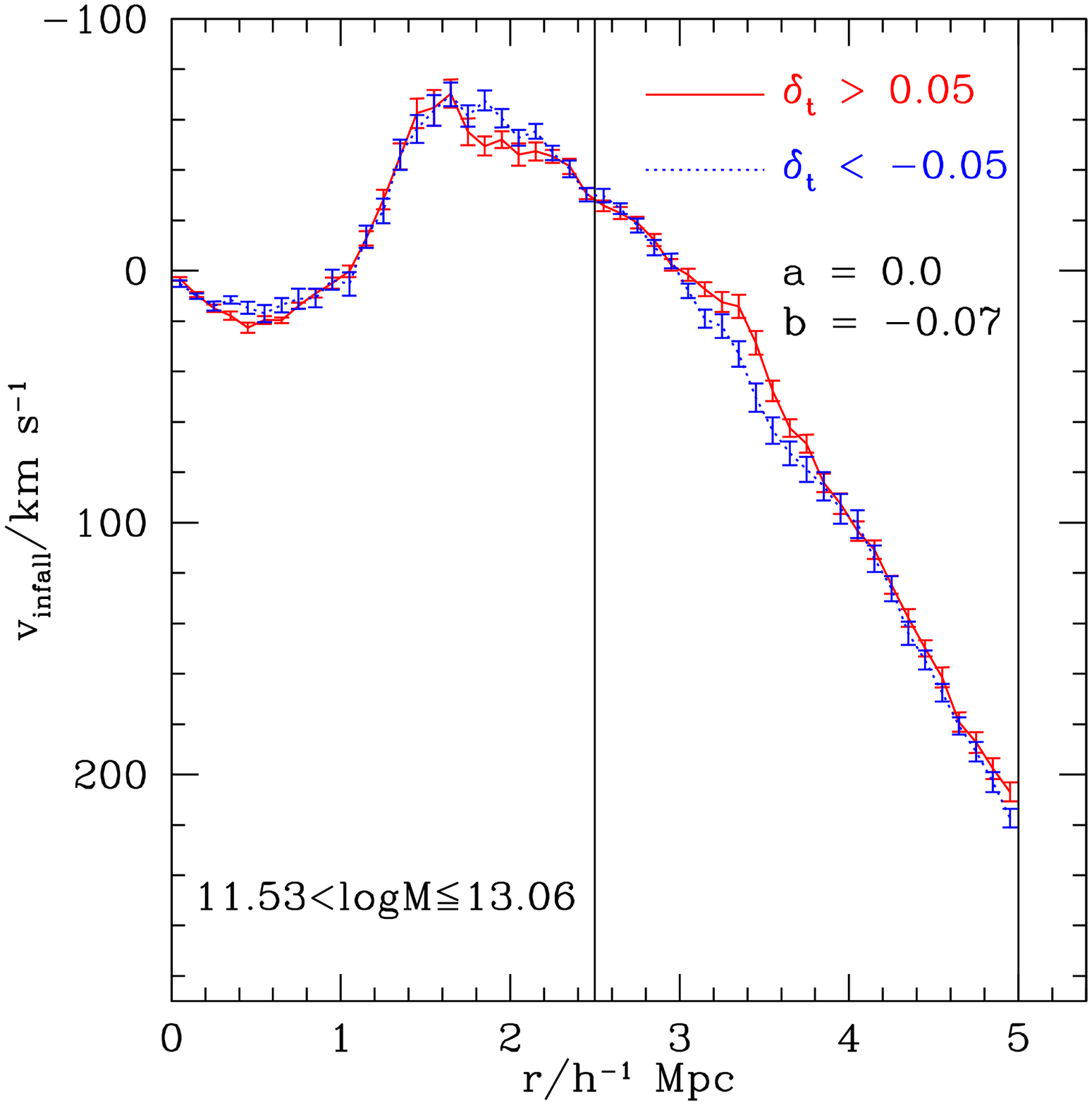} \epsfysize=5.8cm \epsfbox{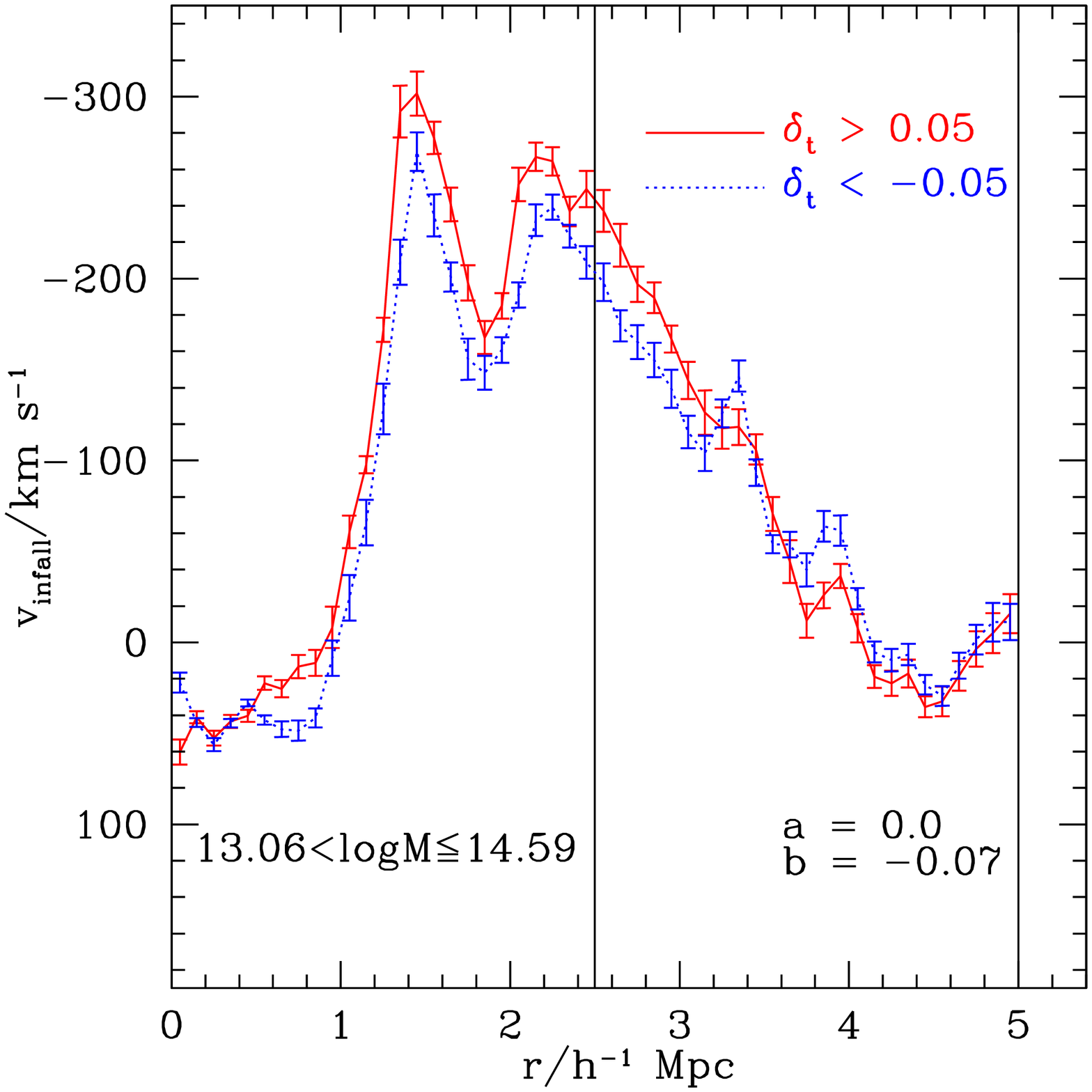}
\caption{Infall velocity profiles for the best-fit parameters
$a = 0$ and $b = -0.07$ 
(see Table \ref{tabla_prm}). Solid lines in red are for old objects and  
dotted lines in blue are for young ones. 
Error bars were calculated using the jackknife method.
Vertical lines mark the range in which the
reduced $\chi^2_{v(r)}$ is calculated. 
The lower-mass bin is on the left-hand panel,
whereas the more massive bin is on the right-hand panel. 
The mass $M$ shown in the figure key is in units of $h^{-1}$ M$_{\odot}$.
Old and young objects
show very similar infall velocity profiles, 
irrespective of the range in mass.
}
\label{best_vel_inf}
\end{figure*}
\begin{figure*}
\leavevmode \epsfysize=5.8cm \epsfbox{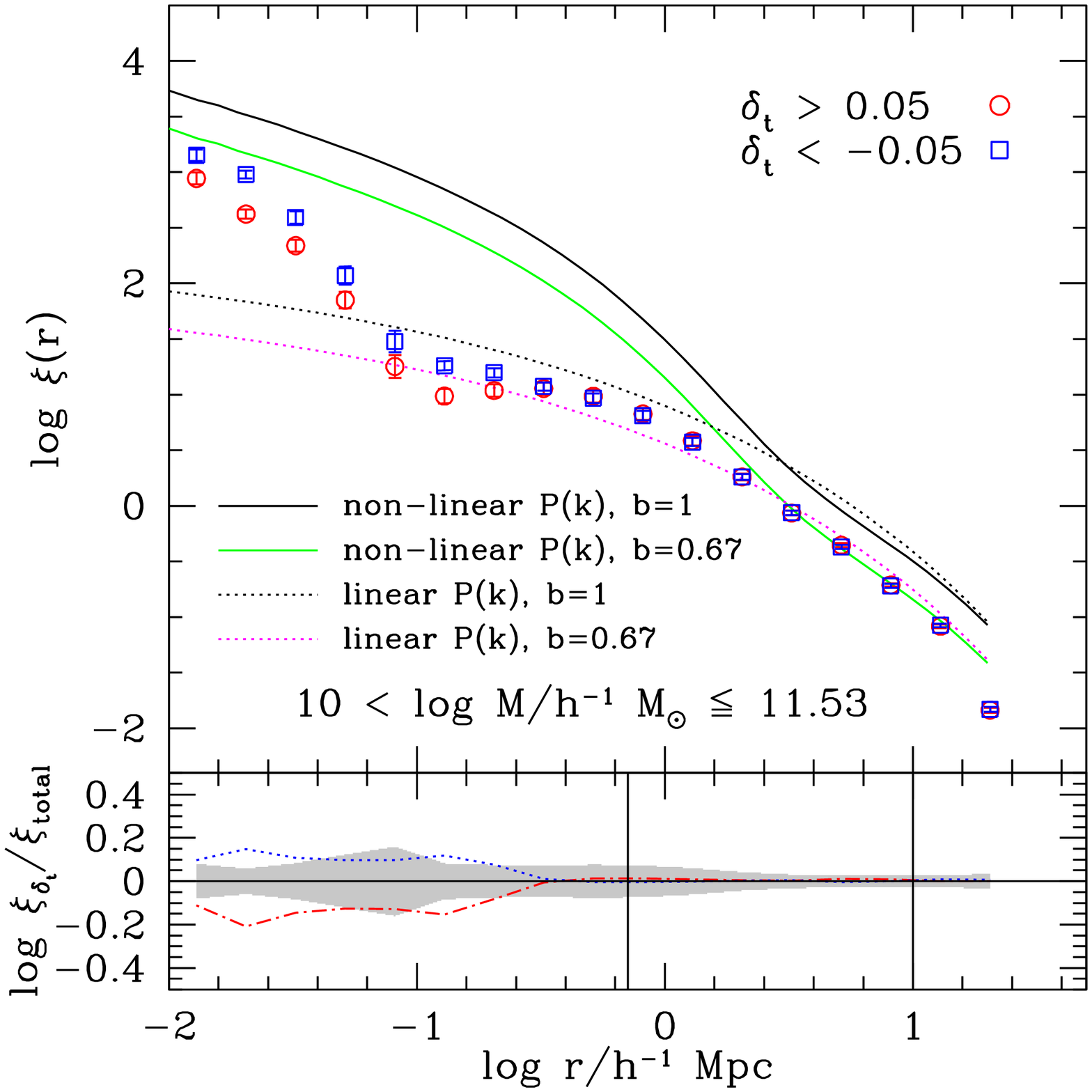} \epsfysize=5.8cm \epsfbox{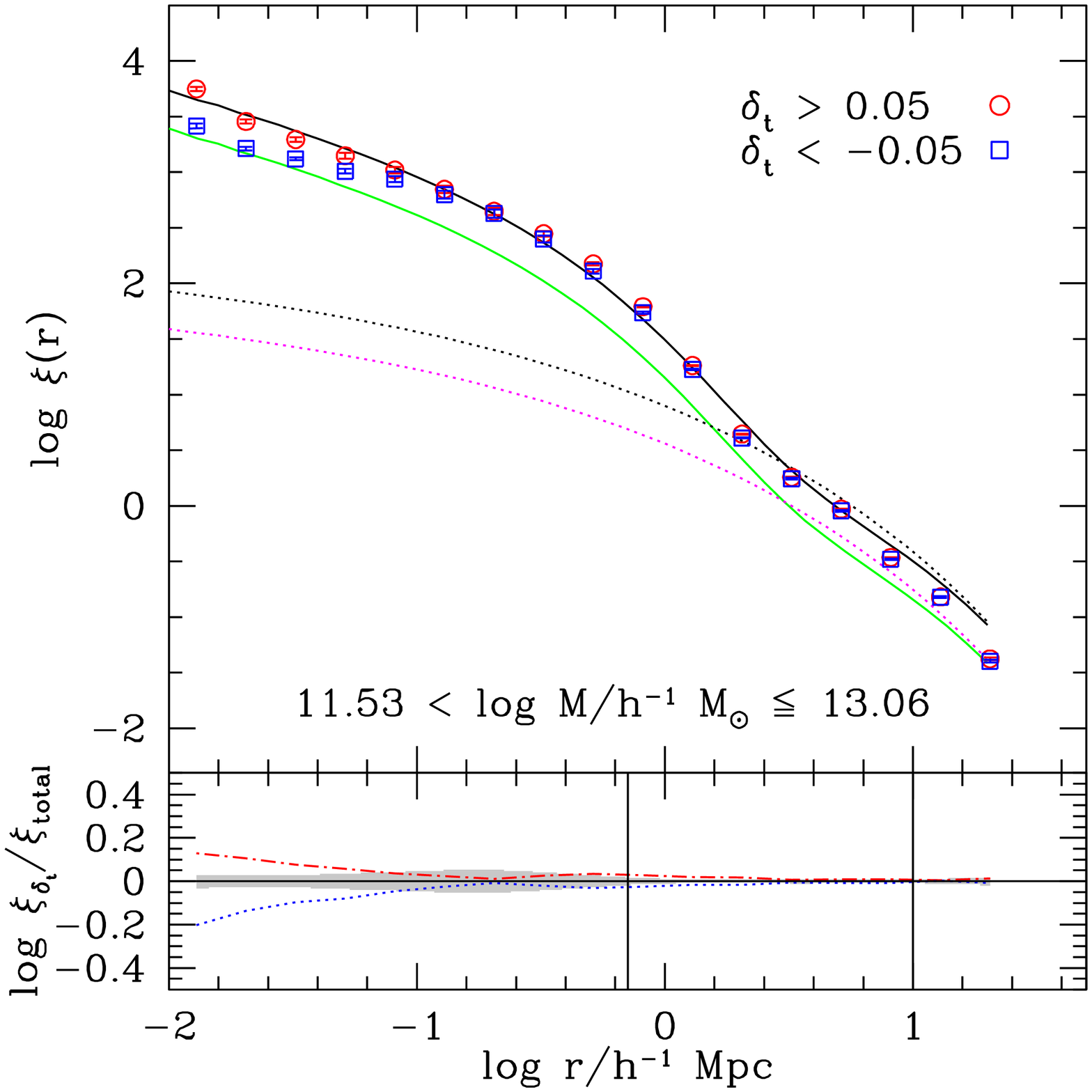} \epsfysize=5.8cm \epsfbox{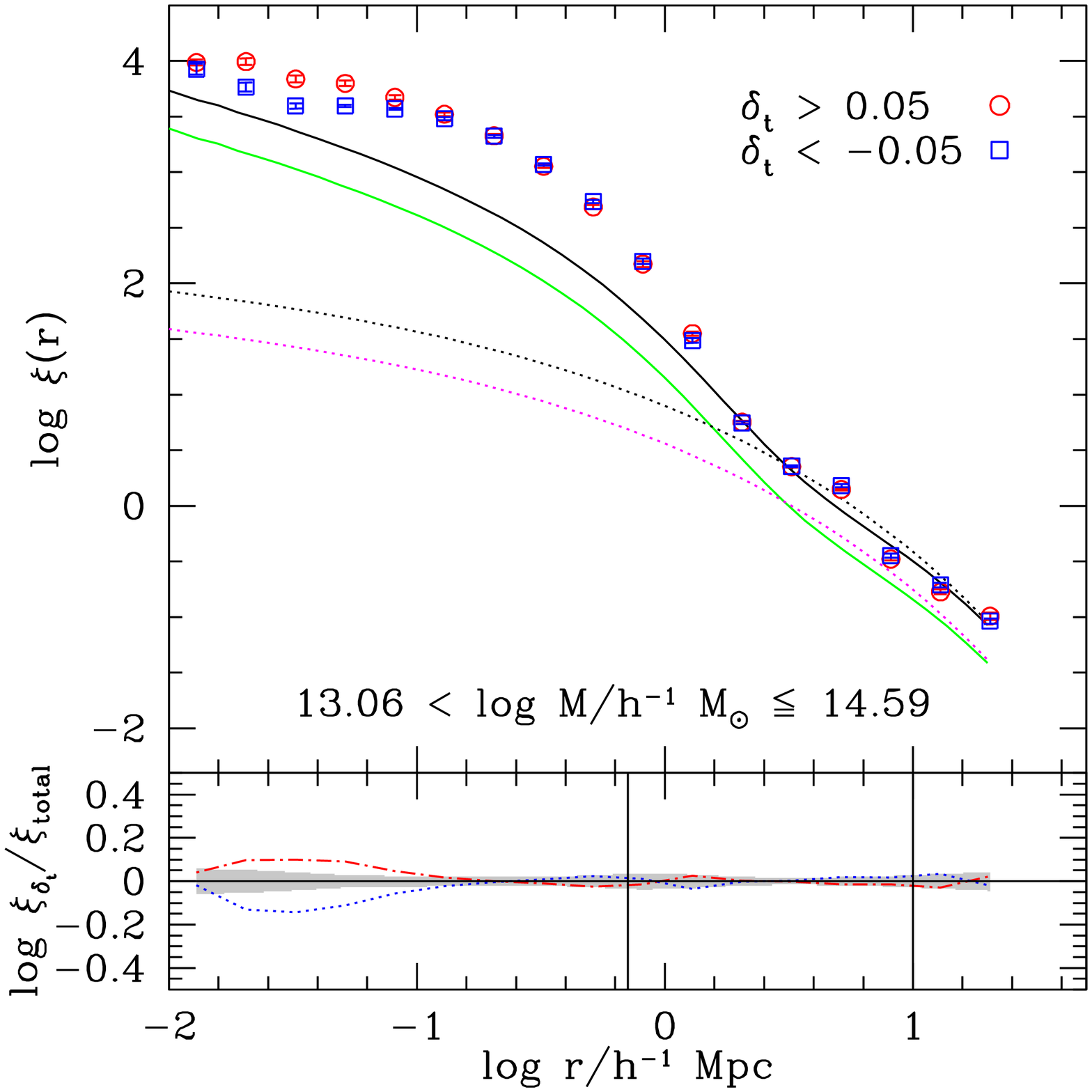}
\caption{Correlation functions for the different mass bins indicated in 
each panel. Old (red circles) and young (blue squares) objects are selected 
by using the radius
parametrisation in Equation (\ref{eq_r_prm}) given by the best-fit parameters
$a = 0$ and $b = -0.07$  (see Table \ref{tabla_prm}).
Error bars were calculated using the jackknife method.
The lines repeated in each top box are obtained from the 
non-linear and linear power spectra, $P(k)$ 
(see Section \ref{section_PS}).
Lower boxes 
are as in Fig. \ref{xi_bines_M}.
The vertical lines mark the range in which the
reduced $\chi^2_{\xi(r)}$ is calculated.
Note that the assembly bias is not present at large scales 
(r $> 1$ $h^{-1}$ Mpc) in any of the mass bins presented. 
For smaller scales, the differences in the clustering 
amplitude between old and young populations are typically
below a factor of two.
}
\label{best_xi}
\end{figure*}

\begin{table}
  \centering
\caption{Best-fit parameters $a$ and $b$ from Equation (\ref{eq_r_prm}).
The last column shows the reduced $\chi^2$ value.}
\begin{tabular}{c c c c c c}\\

\hline
$a$ &   $b$  &  $\chi^2_{v(r)}$ & $\chi^2_{\xi(r)}$ & $\chi^2$ \\   
\hline
\newline

0.00 & -0.07 &  7.80 & 23.68 & 31.48     \\
0.20 & -0.02 &  15.57 & 44.36 & 59.93     \\
\hline
\bigskip
\end{tabular}
\label{tabla_prm}
\end{table}

The best-fit values are shown in  
Table \ref{tabla_prm}. 
They were obtained by marginalising the reduced $\chi^2$ for both
the infall velocity and correlation function statistics. 
Such marginalisation was done by integrating the
likelihood

\begin{equation}
f(\chi) = e^{-(\chi - \chi_{min})^2/2}  \textrm{ ,} 
\end{equation}
\\
where $\chi_{min}$ is the minimum reduced $\chi$ value for a
specific set of parameters. 
The final value
is simply the sum of both results (last column in Table \ref{tabla_prm}).
We find that $f(\chi)$ has two maxima.
The one corresponding to the best fit is that with $a = 0$ and $b = -0.07$.  
The fit corresponding to the second maximum
in likelihood
is that with $a = 0.2$ and $b = -0.02$.
It is worth to mention that the reduced $\chi^2$ value 
allows us to find the
best parameters for a given sample and is not used with the aim of looking for 
the ideal parameters that would result in $\chi^2 \lesssim$ 1, 
since Eq. (\ref{eq_r_prm}) is only intended as an approximation to a more precise peak height proxy.

The infall velocity profiles and correlation functions
for the parameters $a = 0$ and $b = -0.07$ are shown
in Figures \ref{best_vel_inf} and \ref{best_xi}, respectively.
Notice that in this case the size of the sphere in Equation (\ref{eq_r_prm}) depends only 
on the halo mass, specifically $r=-0.07 \log(M_{vir}/M_{nl}$).
The infall velocity profiles for 
old and young galaxies are very similar
for each mass range. Furthermore, the correlation functions for 
these populations 
are remarkably similar at scales r $> 1$ $h^{-1}$ Mpc for each mass bin,
indicating that the assembly bias is not present using this redefinition
of overdensity peak height.  For the case $a = 0.2$ and $b = -0.02$, which depends on both the mass and
age, we obtain similar correlation functions and infall velocity profiles,
although with slight amplitude differences between populations of equal masses but 
different ages (not shown in Figs. \ref{best_vel_inf} and \ref{best_xi}
to improve the clarity of the figures).

Notice that with this new definition of peak height the one-halo terms
of old and young objects of equal mass are comparable, a 
property which the virial mass was not able to produce.

\begin{table*}
  \centering
\caption{Maximum and median radii, $r_{max}$ and $<r>$, respectively, 
in physical units ($h^{-1}$ kpc) 
from Equation (\ref{eq_r_prm}), 
as given by the best-fit parameters in Table \ref{tabla_prm} for all objects and, 
also, split among
the old and young populations of galaxies.
The ranges in virial mass are those shown in Fig. \ref{best_ratio}.
}
\begin{tabular}{c c c c c c c c c}\\

\hline
best-fit params.            &   ages  & $r_{max}$ & & $<r>$ & $r_{max}$ &  & $<r>$   \\   
\hline
& &log($M_{vir} /h^{-1} M_{\odot})$ & = & 10.3 - 10.7 &log($M_{vir} /h^{-1} M_{\odot})$ & = & 10.8 - 11.4 \\
\hline
\newline

$a$ = 0.0, $b= -0.07$ & all                  & 215.5      &     & 205.3              & 180.6     &      &  165.4        \\
                      & old                  & 215.5       &    & 204.2              & 180.6     &      &  166.3        \\
                      & young                & 215.5        &   & 207.5              & 180.6     &      &  163.1  \\\\

$a$ = 0.2, $b= -0.02$ & all                & 415.9          & & 55.9               & 445.2      &     &  86.6    \\               
                      & old                & 415.9          & & 101.8              & 445.2     &       & 123     \\                            
                      & young              & 383.5          & & 51.1               & 407.8    &       &  76.6  \\
\hline
\end{tabular}
\label{tabla_radius}
\end{table*}

\begin{figure}
\leavevmode \epsfysize=8.9cm \epsfbox{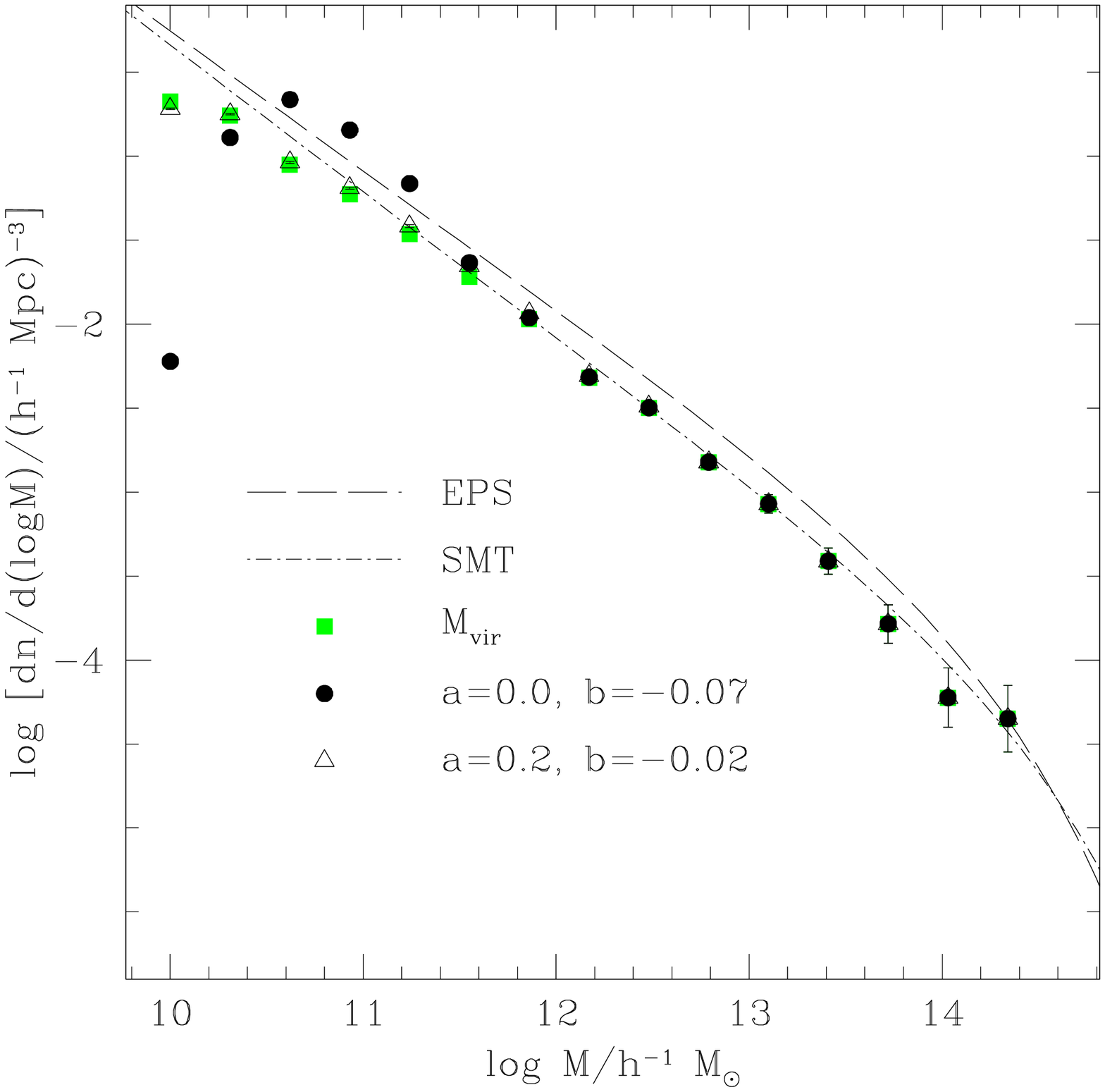}  
\caption{Mass function obtained from using the virial mass 
($M_{vir}$, filled green squares) 
and the new masses from the best-fit parameters $a=0$, $b=-0.07$ 
(filled black circles) and $a=0.2$, $b=-0.02$ (open black triangles).
Error bars correspond to the Poisson error. 
For comparison, we plot the mass functions from the EPS
(long-dashed line) and 
the SMT (dot-dashed line) models.
The best agreement with the SMT mass function is shown by the results from virial masses and 
those from the second-best set of parameters
$a=0.2$ and $b=-0.02$.
}
\label{fig_mass_function}
\end{figure}

Figure \ref{fig_mass_function} shows the mass function for the parameters
$a = 0$ and $b = -0.07$ as filled black circles, for $a=0.2$ and $b=-0.02$ as open triangles,
and for the virial mass as filled green squares. 
For comparison,
the predicted mass functions from the extended Press-Schechter
model (EPS) and from the Sheth, Mo, \& Tormen (2001, SMT) model 
are shown as long-dashed and
dot-dashed lines, respectively. 
At low masses, 
the $a=0$, $b=-0.07$ distribution shows an
unphysical behaviour, with very few objects at 
$M \sim$ 10$^{10}$ $h^{-1}$ M$_{\odot}$
and a bump at $\sim 10^{10.7}$ $h^{-1}$ M$_{\odot}$. 
This means that most of the galaxies hosted 
by haloes in this range
of virial mass ($M_{vir}$, green squares) changed their masses to 
$M \sim$ 10$^{10.7}$ $h^{-1}$ M$_{\odot}$ 
after using these parameters. 
However, the mass function changes only slightly with respect to that of the 
virial mass when using the second-best fit values $a = 0.2$ and $b = -0.02$ 
(open black triangles), which also reduces the assembly bias by introducing a dependence
on the age. None of the two parametrisations change the mass function at 
$M \geq$ 10$^{12}$ $h^{-1} M_{\odot}$, 
and therefore in this range their mass functions and the one resulting from the virial mass 
are all consistent with the SMT prediction.
Therefore, we consider the second-best fit a better candidate since, by
introducing a smaller variation in the mass, we find no assembly bias and a good
agreement with SMT.

Figure \ref{best_ratio} shows the distribution of $r/r_{vir}$ for two
different bins in \emph{virial} mass 
for the two sets of best-fit parameters in Table \ref{tabla_prm}.
Both panels show that most of the galaxies keep their
original halo masses, $M_{vir}$, when using the best-fit parameters
$a = 0.2$ and $b = -0.02$ (solid lines). 
For the lower-mass bin (left panel), these
galaxies have a median value of $r = r_{vir}$.
For the case
$a = 0$ and $b = -0.07$ (dashed lines), 
they have a median value of $r \sim 4 $ $r_{vir}$.
The maximum and median radii (in units of $r_{vir}$)
and, also, the number of 
objects which change their mass,
decrease for higher virial masses. 
Table \ref{tabla_radius} shows these radii in units of kpc.
All the galaxies with 
$M_{vir} \geq 6\times10^{12}$ $h^{-1} M_{\odot}$ 
conserved their virial masses, i.e. $M = M_{vir}$, in both cases.  
This means that
some objects which were initially considered as those with low peak heights,
as given by their virial mass,
are now associated to regions with higher overdensities, 
particularly for 
low virial masses.

\section{Properties of the new peaks}
\label{sec_freq}

\begin{figure*}
\leavevmode \epsfysize=8.5cm \epsfbox{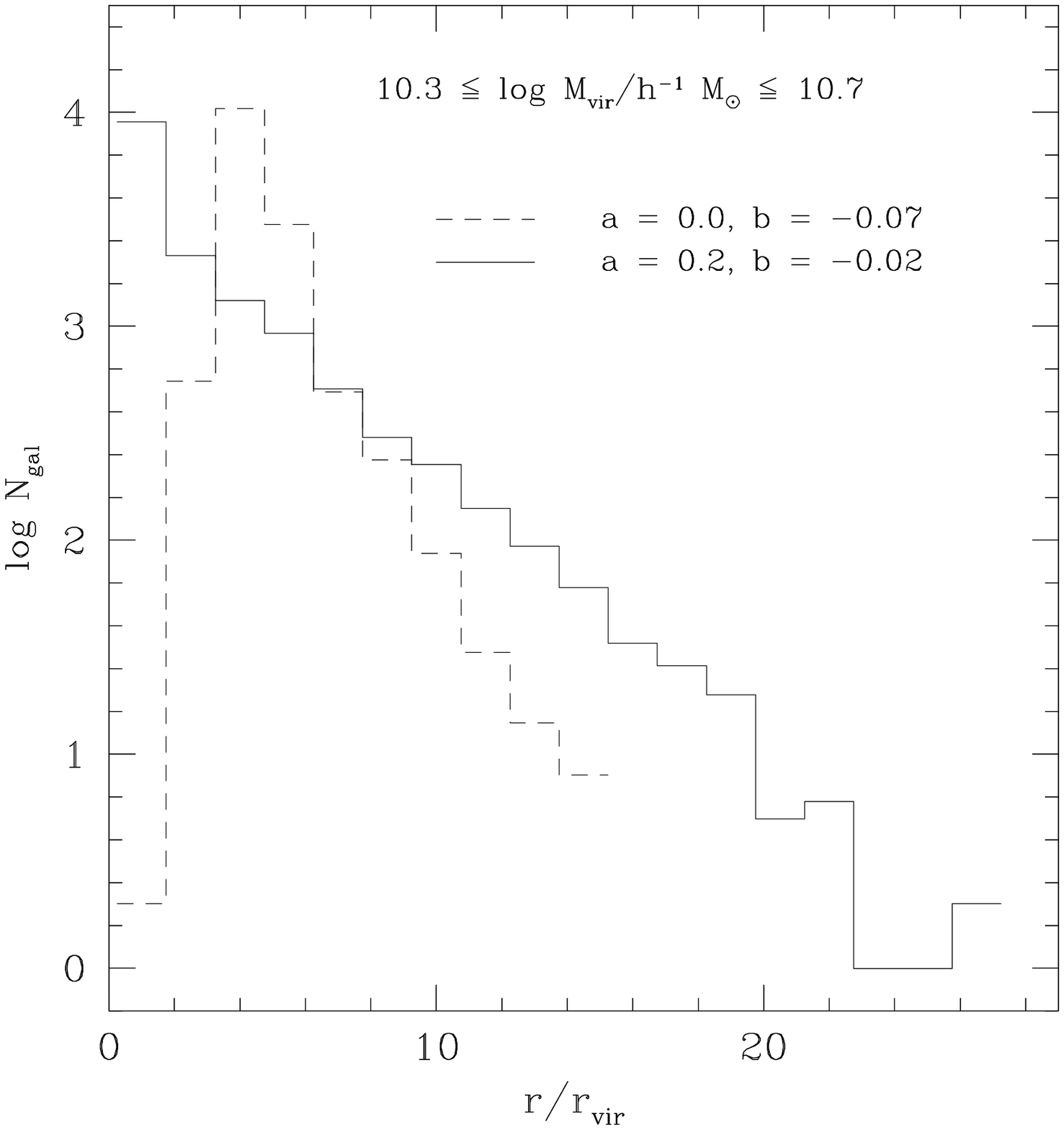} \epsfysize=8.5cm \epsfbox{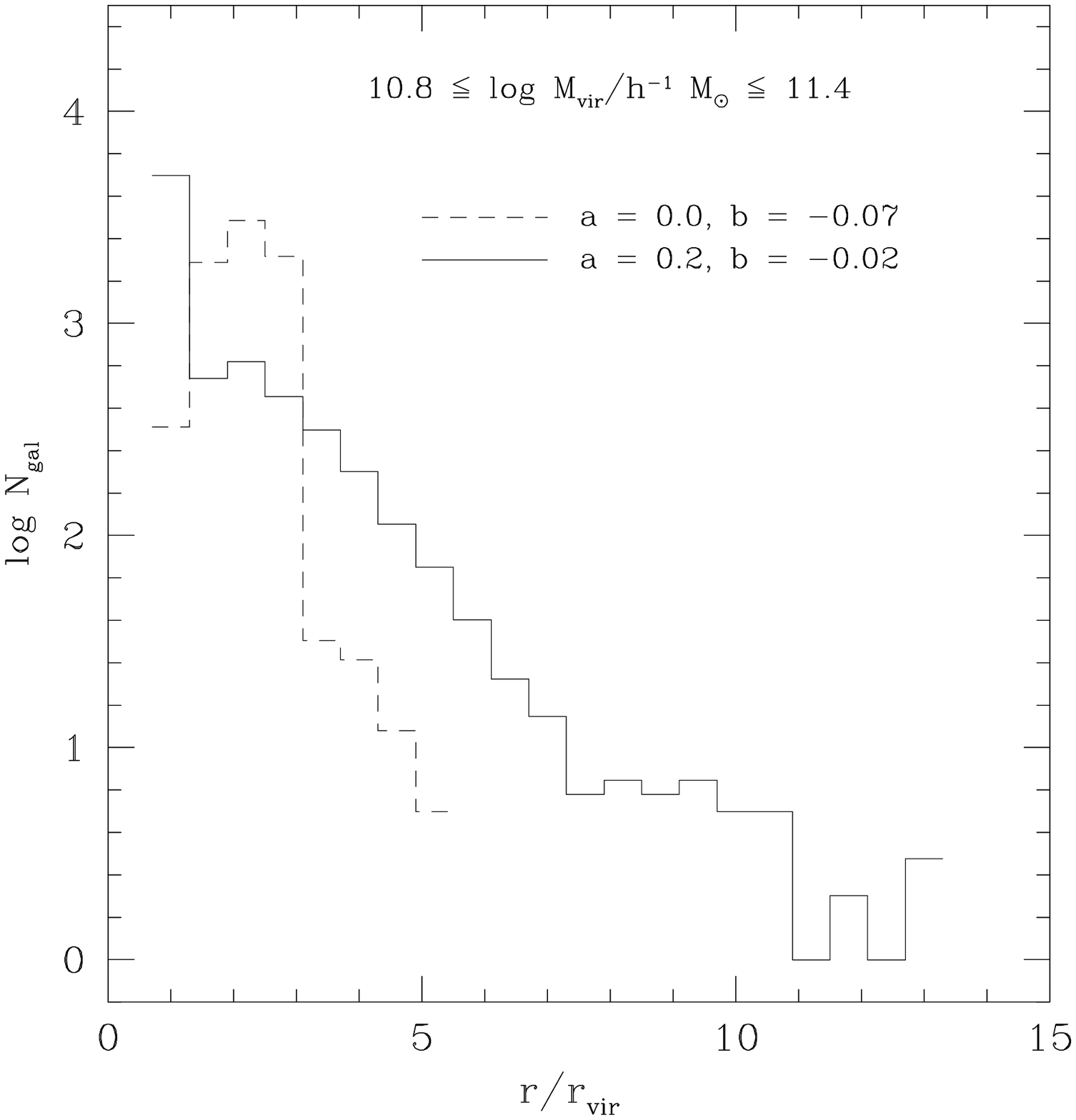} 
\caption{Distribution of $r/r_{vir}$ for different ranges of 
virial mass $M_{vir}$ 
for the best-fit parameters in Table \ref{tabla_prm}. 
For higher values of $M_{vir}$, fewer objects change
their halo masses. 
All the galaxies with log($M_{vir}/h^{-1}$ M$_{\odot}$) $\geq$ 12.8   
conserved their virial masses (i.e. $r = r_{vir}$).  
}
\label{best_ratio}
\end{figure*}

\begin{figure*}
\leavevmode \epsfysize=8.5cm \epsfbox{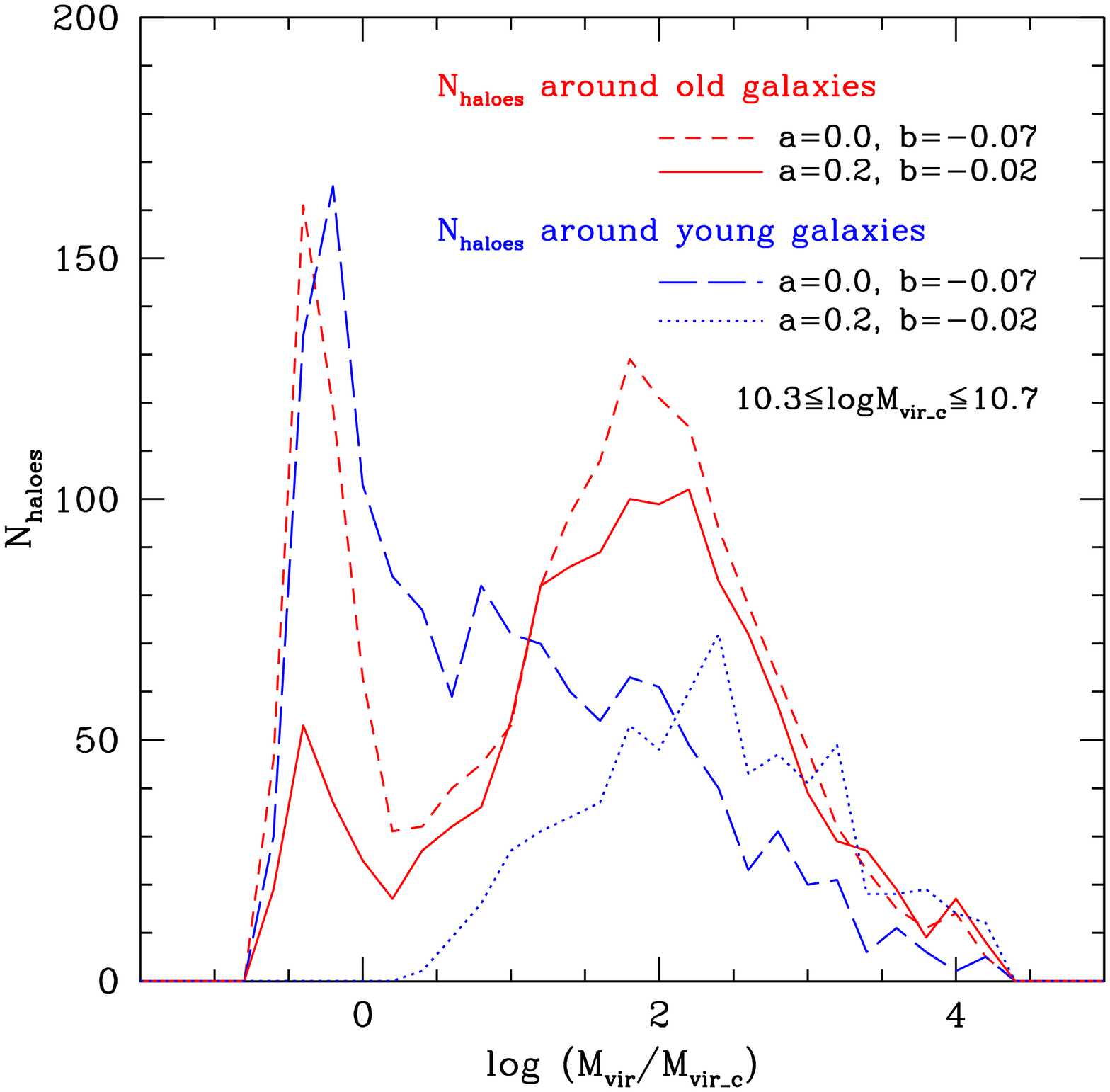}
\leavevmode \epsfysize=8.5cm \epsfbox{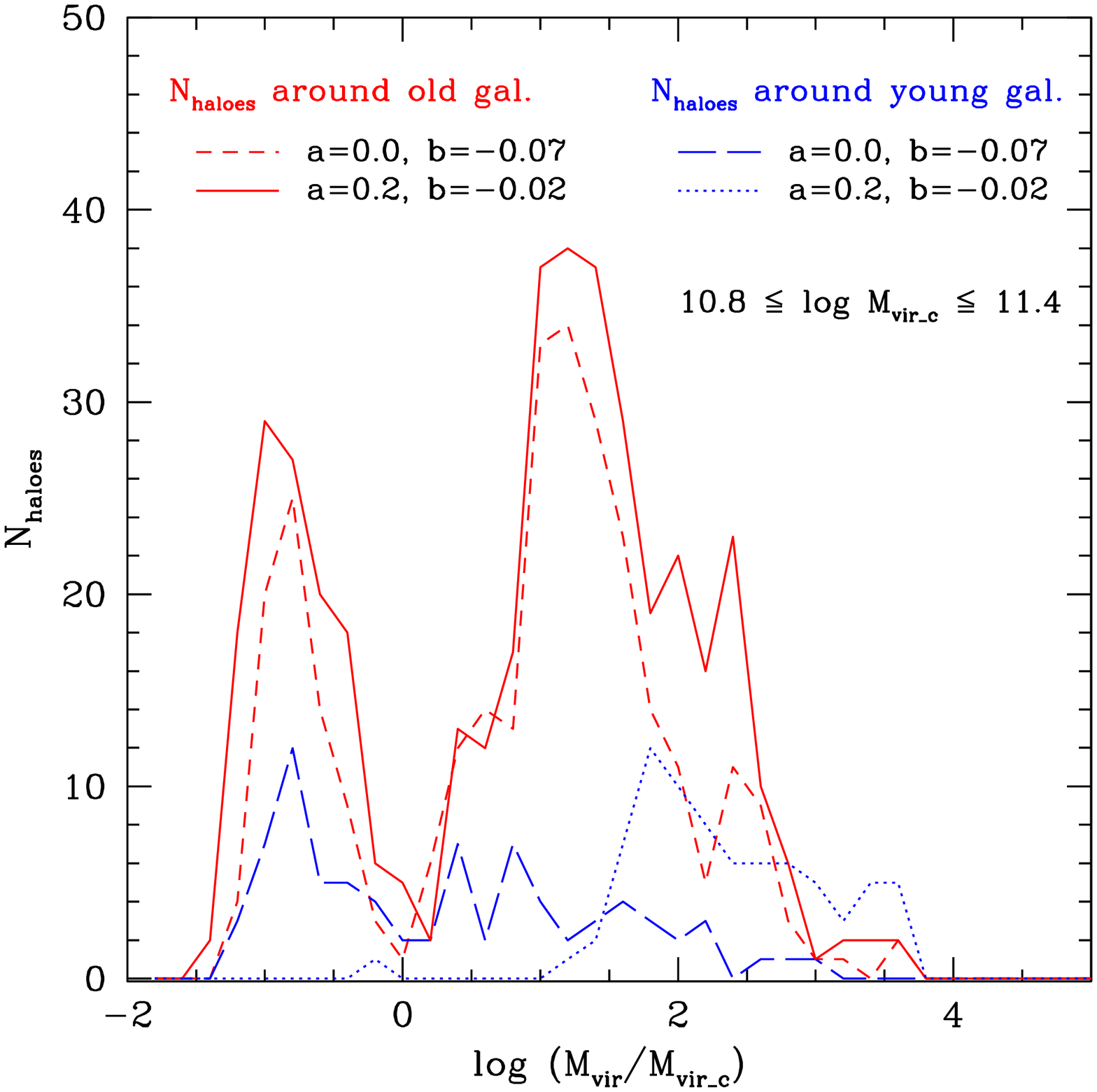}
\caption{Number of haloes inside radius $r$ 
(Eq. \ref{eq_r_prm}) 
given by the best-fit parameters in Table \ref{tabla_prm}:
$a = 0$, $b = -0.07$ (dashed and long-dashed curves around old and young galaxies, 
respectively) and 
$a = 0.2$, $b = -0.02$ (solid and dotted curves around old and young galaxies, respectively).
The results are plotted as a function
of the virial mass 
normalised by the virial mass 
of the central galaxy, $M_{vir\_c}$. The mass range is shown in each panel,
where $M_{vir\_c}$ is in units of $h^{-1}$ M$_{\odot}$.
}
\label{freq}
\end{figure*}

We have presented 
a new 
proxy for the peak height that can account for the assembly bias
at large scales (Section \ref{sec_redefinition_full}). 
In some cases this model considers the mass enclosed by radii greater than
the virial radius,
inside which  
one could be including other haloes. 
In order to see differences between the old and young populations and how these
could affect the statistics for 
$\xi(r)$ and $v_{inf}(r)$, 
Figure \ref{freq} shows
the number of haloes
inside each new peak height (when $r > r_{vir}$),
excluding the central galaxy,
as a function of the ratio between their virial mass  
and the virial mass of the central galaxy, $M_{vir\_c}$,
for the mass ranges and parameters shown in 
Figure \ref{best_ratio}. 
As can be seen from both panels, there is a trend where
the number of haloes contained in spheres of size $r$ 
around young galaxies (blue) is lower than that for spheres  
around old objects (red),  
the effect being  stronger for higher
virial masses (right panel).
Therefore, the peak for 
an old galaxy, after taking into account
the parametrisation of the radius from Equation (\ref{eq_r_prm}), 
adds more haloes and mass than for a young object.
Furthermore, we can see that the higher the virial mass, 
the lower is the influence of other haloes  
in defining the new peak height for galaxies.

Another interesting result is that, 
for the parameters $a = 0.2$ and $b = -0.02$,
both old (solid red lines) and young (dotted blue lines) 
populations tend to add massive peaks, 
as can be seen from the left panel of Figure \ref{freq}, 
which shows that the maximum ratio where the 
distribution is non-zero
is log($M_{vir}/M_{vir\_c}$) $\sim$ 4, but the minimum is 
log($M_{vir}/M_{vir\_c}$) $\sim -0.6$ and 0.4 for the old and young 
galaxies, respectively.
For the case $a = 0$ and $b = -0.07$,
young objects (long-dashed blue curves)
show a peak
around $M_{vir}$ = 0.6$\times$$M_{vir\_c}$,
but they include a broad range of more massive haloes.
Old objects (dashed red curves) are characterised by this 
same behaviour and additionally show a peak at 
log($M_{vir}/M_{vir\_c}$) $\sim$ 1.7.
These results indicate that old, low-mass objects are surrounded preferentially
by high-mass haloes.
The latter is consistent with recent results 
which show that
old, low-mass galaxies suffer truncation of matter by nearby massive haloes
\cite{Wang07, Dalal08, Hahn09}.
However, our results also indicate that there is a population of low-mass objects 
which are 
surrounded by smaller masses. In particular, 
for the $a = 0.2$ and $b = -0.02$ case, 
this is only seen for old objects, regardless of their $M_{vir\_c}$.
It is possible that galaxies with low and high
$M_{vir}/M_{vir\_c}$ ratios correspond to different aspects of the 
assembly bias phenomenology. This, along with studies of the prevalence of this 
bias by varying the concentration, number of satellites, triaxiality, spin, and 
other halo parameters, are the focus of a forthcoming paper 
(\mbox{Lacerna} et al. in preparation).

\section{Conclusions}
\label{conclusiones}

We have presented a new approach to estimate the overdensity peak height
with the aim to understand the assembly bias effect. 
This is a relevant issue that could affect 
the ability of
the next generation 
of galaxy surveys 
to infer accurate cosmological parameters.
Our method consisted in 
redefining the 
overdensity that characterises each galaxy
using the information of its virial mass and the relative age, $\delta_t$; 
this new definition is proposed as a better alternative than the virial mass.
Wang et al. (2007) pointed out that old, low-mass haloes at $z=0$
are associated to higher overdensities in the initial conditions, 
compared to what
would be expected from their final virial masses.  
Instead of searching for the overdensity at high redshifts, 
we try to obtain a measure of the present-day peak height, which in turn
can be tested using large, $z=0$ surveys.
In order to do this, we measure the assembly bias amplitude using two estimators,
the two-point correlation function and the infall velocity profile.  
We find that when using the mass inside spheres of radius $r$ from
Equation (\ref{eq_r_prm}) with the parameters in Table \ref{tabla_prm}, 
galaxies do not show significant differences in the two-halo regime for objects of 
a given mass range but different age.  Furthermore, the dependence on the age is 
reduced in the one-halo term as well; the biggest difference is of a factor of two 
for the lowest mass bin at a separation of $r \sim$ 150 $h^{-1}$ kpc,
which---when using virial masses---becomes a difference 
of two orders of magnitude in the clustering amplitude at the same scale.

The best-fit parameters $a = 0$ and $b = -0.07$ imply that the 
relative age is not 
strictly necessary (see Equation \ref{eq_r_prm}) 
to find a peak height that includes the mass that has not 
collapsed completely
onto haloes yet, and at the same time traces the assembly bias. 
We found that the best parameters are those that yield median sphere radii 
in the
range of 1 - 4 $r_{vir}$. Clearly, environmental effects out to these distances
are playing the main role in shaping 
the two-halo term, as shown in Fig. \ref{best_ratio}. It is worth to point out
that only low-mass objects, 
with $M_{vir} \leq 6\times10^{12}$ $h^{-1} M_{\odot}$, are 
subject to a change in their peak heights,
which coincides with the mass limit for assembly bias found
by several authors (e.g. Gao et al. 2005).
This is also the case for our second-best fitting parameters,
$a=0.2$ and $b=-0.02$,
which introduce a dependence of the peak height on the age, and help trace the
assembly bias while at the same time produce final masses that are in excellent
agreement with the SMT mass function.  Therefore, this option is the
preferred one to obtain a proxy for the peak height which is not subject
to the assembly bias effect.

Neighbouring massive haloes 
that are typically at distances out to 4 $r_{vir}$ (see Figs. \ref{best_ratio} and \ref{freq})
are probably responsible for these effects. 
These could disrupt the normal growth of small objects and, therefore, affect their ages.
However, we also find a population of haloes which, with the new 
definition, includes nearby low-mass haloes, particularly for old objects.

To summarise, we stress the apparent fact that
particularly for low-mass objects, the \emph{virial} mass
is not an adequate proxy for peak height in the standard EPS picture, because equal 
virial mass objects can actually belong to initial density peaks
of very different amplitude, as evidenced in the large differences shown in the 
2-halo regime by statistics such as the correlation function and infall velocities. 
It is necessary to include a more global environmental component,
i.e. the mass of the region that effectively characterises the peak height.
When the latter is taken into account, we obtain the general prescription where the 
bias responds to the height of the mass peak alone at large scales.  Further work is required
in order to confirm that this proposed parametrisation of the peak height is enough to
account for other variations of clustering of equal mass haloes with
different properties such as concentration, spin, etc.  The next papers in this series
will study this, along with an application of this method to large surveys.

\bigskip
\bigskip

We would like to thank Darren Croton, Diego Garc\'ia Lambas, 
Raul Angulo,
Simon D. M. White, Uros Seljak, and the anonymous referee for useful comments and discussions.
We also thank Marcio Catelan for helpful and detailed comments 
on the manuscript.
We acknowledge support from FONDAP ``Centro de Astrof\'\i sica" $15010003$, BASAL-CATA,
Fondecyt grant No. 1071006, CONICYT, and MECESUP.  IL thanks travel support to attend 
international conferences from Fondo ALMA-CONICYT 31070007 and VRAID.

\label{lastpage}


\begin{thebibliography}{}


 \bibitem[Balogh et al. 2004]{Balogh04} 
   Balogh M., et al., 2004, MNRAS, 348, 1355 

 \bibitem[Baugh et al. 2005]{Baugh05}
   Baugh C. M., et al., 2005, MNRAS, 356, 1191 

 \bibitem[Berlind et al. 2003]{Berlind03} 
   Berlind A., et al., 2003, ApJ, 593, 1

 \bibitem[Bett et al. 2007]{Bett07} 
   Bett P., et al., 2007, MNRAS, 376, 215

 \bibitem[Bond et al. 1991]{Bond91} 
   Bond J. R., Cole S., Efstathiou G., Kaiser N., 1991, ApJ, 379, 440

 \bibitem[Bornancini et al. 2006]{Bornancini06} 
   Bornancini C., Padilla N., Lambas D. G., De Breuck C., 2006, MNRAS, 368, 619

 \bibitem[Boylan-Kolchin et al. 2009]{Boylan-Kolchin09}  
   Boylan-Kolchin M., Springel V., White S. D. M., Jenkins A., Lemson G., 
   2009, MNRAS, 398, 1150

 \bibitem[Ceccarelli et al. 2008]{Cecca08} 
   Ceccarelli L., Padilla N., Lambas D. G., 2008, MNRAS, 390, L9

 \bibitem[Cooper et al. 2010]{Cooper10} 
   Cooper M. C., Gallazzi A., Newman J. A., Yan R., 2010, MNRAS, 402, 1942

 \bibitem[Croton et al. 2007]{Croton07}
   Croton D. J., Gao L., White S. D. M., 2007, MNRAS, 374, 1303

 \bibitem[Dalal et al. 2008]{Dalal08}
   Dalal N., White M., Bond J. R., Shirokov A., 2008, ApJ, 687, 12

 \bibitem[Faltenbacher $\&$ White 2010]{FW10}
   Faltenbacher A., White S. D. M., 2010, ApJ, 708, 469

 \bibitem[Gallazzi et al. 2005]{Gallazzi05}
   Gallazzi A., Charlot S., Brinchmann J., White S. D. M., Tremonti C. A., 
   2005, MNRAS, 362, 41

 \bibitem[Gao et al. 2005]{Gao05} 
   Gao L., Springel V., White S. D. M., 2005, MNRAS, 363, 66

 \bibitem[Gao $\&$ White 2007]{Gao-White07} 
   Gao L., White S. D. M., 2007, MNRAS, 377, L5
  
 \bibitem[Gomez et al. (2003)]{Gomez03} 
   Gomez P. L., et al., 2003, ApJ, 584, 210

 \bibitem[Gonzalez $\&$ Padilla 2009]{Rob-Nelson09} 
   Gonzalez R. E., Padilla N., 2009, MNRAS, 397, 1498

 \bibitem[Gunn $\&$ Gott 1972]{GG72} 
   Gunn J. E., Gott J. R., 1972, ApJ, 176, 1

 \bibitem[Hahn et al. 2009]{Hahn09}
   Hahn O., Porciani C., Dekel A., Carollo C. M., 2009, MNRAS, 398, 1742

 \bibitem[Hester $\&$  Tasitsiomi 2010]{HT10} 
   Hester J. A., Tasitsiomi A., 2010, ApJ, 715, 342

 \bibitem[Kauffmann et al. 1997]{K97} 
   Kauffmann G., Nusser A., Steinmetz M., 1997, \mnras, 286, 795

 \bibitem[Kauffmann et al. 2003]{K03} 
   Kauffmann G., et al., 2003, MNRAS, 341, 33

 \bibitem[Lacey $\&$ Cole (1993)]{Lacey-Cole93}
   Lacey C., Cole S., 1993, MNRAS, 262, 627

 \bibitem[Lagos, Cora, $\&$ Padilla 2008]{Lagos08}
   Lagos C., Cora S. A., Padilla N., 2008, MNRAS, 388, 587

 \bibitem[Lagos, Cora, $\&$ Padilla 2009]{Lagos09}
   Lagos C., Padilla N., Cora S., 2009, MNRAS, 395, 625

 \bibitem[Li et al. 2008]{Li08}
   Li Y., Mo H. J., Gao L., 2008, MNRAS, 389, 1419

 \bibitem[Ludlow et al. 2009]{Ludlow09}
   Ludlow A. D., et al., 2009, ApJ, 692, 931

 \bibitem[Mo $\&$ White 1996]{Mo-White96}
   Mo H. J., White S. D. M., 1996, MNRAS, 282, 347

 \bibitem[Padilla, Garc\'\i a \& Gonz\'alez, 2010]{Padilla10}
   Padilla N. D., Lambas D. G., Gonzalez R., 2010, MNRAS, tmp, 1417

 \bibitem[Percival et al. 2003]{Percival03}
   Percival W. J., Scott D., Peacock J. A., Dunlop J., 2003, MNRAS, 338, L31

 \bibitem[Press $\&$ Schechter 1974]{PS74}
   Press W. H., Schechter P., 1974, ApJ, 187, 425

 \bibitem[Seljak $\&$ Warren 2004]{SW04}
   Seljak U., Warren M. S., 2004, MNRAS, 355, 129

 \bibitem[Sheth, Mo, and Tormen 2001]{SMT01}
   Sheth R. K., Mo H. J., Tormen G., 2001, MNRAS, 323, 1

 \bibitem[Smith et al. 2003]{Smith03}
   Smith R. E., et al., 2003, MNRAS, 341, 1311

 \bibitem[Springel et al. 2005]{Springel05}
   Springel et al., 2005, Nature, 435, 629

 \bibitem[Yang et al. 2003]{Yang03}
   Yang, X., Mo, H. J., van den Bosch, F. C., 2003, MNRAS, 339, 1057

 \bibitem[Wang et al. 2007]{Wang07}
   Wang H., Mo H. J., Jing Y. P., 2007, MNRAS, 375, 633

 \bibitem[Wang et al. 2009]{Wang09}
   Wang H., Mo H. J., Jing Y. P., 2009, MNRAS, 396, 2249

 \bibitem[Wang et al. 2008]{WangYu08}
   Wang Y., Yang X., Mo H. J., van den Bosch F., Weinmann S., Chu Y., 2008, ApJ, 687, 919

 \bibitem[Wechsler et al. 2006]{Wechsler06}
   Wechsler R. H., Zentner A. R., Bullock J. S., Kravtsov A. V., Allgood B., 2006, ApJ, 652, 71

 \bibitem[Wu et al. 2008]{Wu08}
   Wu H., Rozo E., Wechsler R. H., 2008, ApJ, 688, 729

 \bibitem[Zapata et al. 2009]{Zapata09}
   Zapata T., Perez J., Padilla N., Tissera P., 2009, \mnras, 394, 2229

 \bibitem[Zentner 2007]{Zentner07}
   Zentner A. R., 2007, IJMPD, 16, 763

 \bibitem[Zhu et al. 2006]{Zhu06}
   Zhu G., et al., 2006, ApJ, 639, L5

\end{thebibliography}
\end{document}